\documentclass[twocolumn]{revtex4}
\usepackage{amssymb}
\usepackage{latexsym}
\usepackage{epsfig}
\usepackage{amsmath}
\usepackage{graphicx}
\usepackage{epsfig}
\usepackage{bm}
\usepackage{cancel}
\usepackage{amssymb}
\usepackage{float}
\usepackage{subfigure}
\usepackage{dcolumn}
\usepackage[colorlinks]{hyperref}
\usepackage[usenames,dvipsnames]{color}

\hypersetup{
     breaklinks=true,
    pdfstartview={FitH},    
    colorlinks=true,       
    linkcolor=blue,          
    citecolor=red,        
    filecolor=magenta,      
    urlcolor=blue,           
    anchorcolor=green,      
    linktocpage=true
}
\begin{document}
\title{A Generalized Interacting Tsallis Holographic Dark Energy Model and its thermodynamic implications}
\author{Abdulla Al Mamon}
\email{abdulla.physics@gmail.com}
\affiliation{Department of Physics, Vivekananda Satavarshiki Mahavidyalaya (affiliated
to the Vidyasagar University), Manikpara-721513, West Bengal, India}
\author{Amir Hadi Ziaie}
\email{ah.ziaie@maragheh.ac.ir}
\affiliation{Research Institute for Astronomy and Astrophysics of Maragha (RIAAM), University of Maragheh, P.O. Box 55136-553, Maragheh, Iran}
\author{Kazuharu Bamba}
\email{bamba@sss.fukushima-u.ac.jp}
\affiliation{Division of Human Support System, Faculty of Symbiotic
Systems Science, Fukushima University, Fukushima 960-1296, Japan}
\begin{abstract}
The present paper deals with a theoretical model for interacting  Tsallis
holographic dark energy (THDE) whose infrared (IR) cut-off scale is set by the Hubble length. The interaction $Q$ between the dark sectors (dark energy and pressureless dark matter) of the universe has been assumed to be non-gravitational in nature. The functional form of $Q$ is chosen in such a way that it reproduces well known and most used interactions as special cases. We then study the nature of the THDE density parameter, the equation of state parameter, the deceleration parameter and the jerk parameter for this interacting THDE model. Our study shows that the universe exhibits the usual thermal history, namely the successive sequence of radiation, dark matter and dark energy epochs, before resulting in a complete dark energy domination in the far future. It is shown the evolution of the Hubble parameter for our model and compared that with the latest Hubble parameter data. Finally, we also investigate  both the stability and thermodynamic nature of this model in the present context.
\end{abstract}
\maketitle
Keywords: Tsallis holographic dark energy, interaction, thermodynamics.
\section{Introduction}
Many cosmological observations indicate that our Universe is now experiencing an accelerated expansion phase \cite{acc1,acc2,acc3,acc4,acc5}. A possible candidate to explain this cosmic acceleration is to consider some exotic matter, dubbed as dark energy (DE) which consists of approximately 68\% of the total energy budget of our universe. However, the origin and nature of this DE are absolutely unknown. On the other hand, the second largest component of our universe is the dark matter (DM) which takes around 28\% of the total energy density of the universe. Like the DE sector, DM sector is also not very well understood. Till now, a large number of theoretical models are taken into account to accommodate the present phase of acceleration and some excellent reviews on this topic can be found in \cite{de1,de2,de3}. However, the problem of the onset and nature of cosmic acceleration remains an open challenge of modern cosmology at present.\\
\par In this context, holographic dark
energy (HDE) is an interesting attempt to solve this problem (for details, see \cite{Hooft,Hooft1995,Cohen}) and some of its various scenarios can be found in
\cite{hde1,hde2,hde3,hde4,hde5,hde6,hde7,hde8,hde9,hde10,hde11,hde12,hde13,hde14,hde15,hde16,hde17,hde18,hde19,hde20,hde21}. In particular, a new HDE model has been proposed by using the holographic hypothesis and the Tsallis entropy \cite{tsahde}, named Tsallis holographic dark energy (THDE) \cite{THDE,tnote}. As a result, recently, several THDE models have been investigated and explored in different scenarios with an aim to search for the dynamics of the universe and one can look into \cite{thde1,thde2,thde3,thde4,thde5,thde6,thde7,thde8,thde9,thde10,thde11,thde12d,thde13n} for a comprehensive review. \\ 
\par It is important to mention that observations admit an interaction between the dark sectors (DM and DE) of cosmos which can solve the coincidence problem and the tension in current observational values of the Hubble parameter \cite{obsint01,obsint02,obsint03,gdo1,gdo2,ob1,ob2,ob3,ob4,ob5,ob6,ob7,h1,ig1,ig2,ig3,ig4,id1,im1,im2,ie1,iaamnc}. The scenario of interaction between DM and DE is one of such alternative models, which is the main subject interest of the present work. Recently, Zadeh et al. \cite{tnote} investigated the evolution of the THDE models with different IR cutoffs and studied their cosmological consequences under the assumption of a mutual interaction between the dark sectors of the universe. Following \cite{tnote}, in this work, we are also interested in studying the dynamics of a flat FRW universe filled with a pressureless matter and THDE in an interacting scenario. In particular, we explore consequences of interacting THDE model in a more general scenario. In our setups, we study the evolution of our universe by considering an interaction between DM and THDE whose IR cutoff is the Hubble horizon. The nature of the THDE density parameter, the deceleration parameter, the jerk parameter and the THDE equation of state parameter have also been studied for the present model. Furthermore, we also investigate the stability and thermodynamic nature of this particular model in the present scenario. However, the present work is more general and also different from the similar work by Zadeh et al. \cite{tnote} in different ways. Firstly, in this paper, the functional form of the interaction term is chosen in such a way that it can reproduce some well known and most used interactions (including \cite{tnote}) in the literature for some special cases (for details, see section \ref{sec2}). Secondly, we study the evolution of jerk parameter for this general interaction term. We also plot the Hubble parameter for our model and compared that with the latest Hubble parameter data. Additionally, we go one step further by investigating this scenario taking into account the thermodynamic considerations. In particular, we study the nature of the total entropy of the universe surrounded by a cosmological horizon. For completeness, we extend the interacting THDE model, in the case where the radiation fluid is also present.\\
\par The paper is organized as follows. In the next section, we present a THDE model with Hubble scale as IR cutoff. Additionally, the results of considering a mutual interaction between the dark sectors of the universe are also investigated. In section \ref{sec-thermo}, we also explore the thermodynamical properties of the present model.  Finally, in section \ref{conclusion}, we summarize the conclusions of this
work.\\
\par Throughout the text, the symbol dot indicates derivative with respect to the cosmic time and a subscript zero refers to any quantity calculated at the present time.
\section{Interacting THDE with Hubble Cutoff}\label{sec2}
The THDE model is based on the modified entropy-area relation \cite{tsahde} and the holographic dark energy hypothesis, was proposed in \cite{THDE} by introducing the following energy density
\begin{eqnarray}\label{Trho}
\rho_D=BL^{2\delta-4}
\end{eqnarray}
where $B$ is an unknown parameter and $\delta$ denotes the non-
additivity parameter \cite{tsahde,THDE}. It is worth mentioning that in the special case $\delta=1$ the above relation gives the usual HDE $\rho_D=BL^{-2}$, with $B=3c^{2}m^{2}_{p}$, and $c^2$ and $m_p$ are the model parameter and the reduced Planck mass, respectively. Additionally, for $\delta=2$ equation (\ref{Trho}) gives the standard cosmological constant ($\Lambda$) case $\rho_D=\Lambda=$ constant. In this way THDE is indeed a more general framework than the standard HDE and hereafter,   we focus on the general case, i.e., $\delta \neq 1$ and $\delta \neq 2$. At this stage, since the positive body of DE itself is not understood yet, as a possible approach to acquire the clue to reveal the nature of DE, we propose an application of holography and entropy relations to the whole universe, which is a gravitational and non-extensive system. In this work, especially, we consider the Tsallis entropy and we use the generalized definition of the universe horizon entropy, given in equation (\ref{Trho}). If the physical mechanism of holography and the fundamental relation between gravitation and thermodynamics are found in the future studies, it is expected that the Lagrangian for HDE can be written explicitly and we can derive the equations of motion for such a physical system leading to the energy density of DE component in equation (\ref{Trho}). Although we do not have a form of the Lagrangian, through a phenomenological approach, we can acquire the resultant representation of the DE density. This is the positive motivation and a kind of justification to investigate HDE models. \\
\par By considering the Hubble horizon as the IR cutoff, i.e., $L=H^{-1}$, the energy density corresponding to THDE is obtained as
\begin{eqnarray}\label{Hrho}
\rho_D=BH^{-2\delta+4},
\end{eqnarray} 
In the large scale, our universe is homogeneous and isotropic and its geometry is best described by the spatially flat Friedmann-Robertson-Walker (FRW) metric
\begin{eqnarray}\label{frw}
ds^{2}=dt^{2}-a^{2}\left( t\right) \left[ dr^2
+r^{2}d\Omega^{2}\right],
\end{eqnarray}
where $a(t)$ is the scale factor of the universe. Now, in such a spacetime, one can write down Friedmann equations as \cite{de1}
\begin{eqnarray}\label{frd}
H^{2}=\frac{1}{3m_{p}^{2}}\left(\rho_{m}+\rho_{D}\right)
\end{eqnarray}
where, $H=\frac{\dot{a}}{a}$ is the Hubble parameter and an overhead dot represents derivative with respect to the cosmic time $t$. Also, $\rho_m$ and $\rho_D$ represent the energy density of pressureless matter and the THDE density, respectively. The energy density parameter of THDE and pressureless matter can be expressed as 
\begin{eqnarray}\label{3}
&&\Omega_{D}=\frac{\rho_{D}}{\rho_c}=\frac{B}{3m_{p}^{2}}H^{-2\delta+2}\\
&&\Omega_{m}=\frac{\rho_{m}}{\rho_c}
\end{eqnarray}
where, $\rho_c=3m_{p}^{2}H^{2}$ denotes the critical energy
density. Now, equation (\ref{frd}) can be written as
\begin{eqnarray}\label{u}
\Omega_{m}+\Omega_{D}=1
\end{eqnarray}
Also, the ratio of the energy densities is obtained as
\begin{equation}\label{r2}
r=\frac{\rho_m}{\rho_D}=\frac{\Omega_m}{\Omega_D}
\end{equation}
Moreover, we assume that DM and THDE interact with each other. Accordingly, the energy conservation equations become
\begin{eqnarray}\label{conm}
&&\dot{\rho}_m+3H\rho_m=Q\\
&&\dot{\rho}_D+3H(1+\omega_D)\rho_D=-Q\label{conD}
\end{eqnarray}
where $\omega_D\equiv \frac{p_D}{\rho_D}$ denotes the {\it equation of state} (EoS) parameter of THDE, $p_{D}$ is the pressure of THDE and $Q$ indicates the rate of energy exchange between the dark sectors (DM and THDE). Positive value of $Q$ indicates that there is an energy transfer from the THDE to the DM, while for $Q<0$, the reverse scenario happens. On the other hand, if $Q=0$ (i.e., non-interacting case), then the DM evolve as, $\rho_m \propto a^{-3}$. Hence, the interaction between the dark sectors is indeed a more general scenario to unveil the dynamics of the universe. In fact, there are many proposed interactions in the literature to study the dynamics of the universe (for details, one can look into \cite{ig1,ig2,ig3,ig4,id1,im1,im2,ie1,iaamnc,HDESChange} and references therein), however, the exact functional form of  $Q$ is still unknown to us. From the continuity equations (\ref{conm}) and (\ref{conD}), one can see that the interaction $Q$ could be any arbitrary function of the parameters $\rho_m$, $\rho_D$ and $H$. So, naturally, one can construct various interacting models to understand the dynamics of the universe in this framework. For mathematical simplicity, in the present work, we assume that the interaction is a linear combination of the dark sector densities given as 
\begin{equation}\label{genQ}
Q=3H(b^2_{1}\rho_{m} + b^2_{2}\rho_{D}),
\end{equation}
where, the parameters $b_{1}$ and $b_{2}$ are dimensionless constants. This type of functional forms of $Q$ has been studied recently by several authors \cite{ig1,ig2,ig3,ig4,id1} and the particular cases $b_{1}=0$, $b^2_{2}=\lambda$ in Ref. \cite{id1}, $b_{2}=0$, $b^2_{1}=\alpha$ in Refs. \cite{im1,im2},  $b^2_{1}=\frac{\lambda_m}{3}$, $b^2_{2}=\frac{\lambda_D}{3}$ in Ref. \cite{ig1}, $b_{1}=b_{2}=b$ in Ref. \cite{tnote} and $b_{2}=0$, $b_{1}=b$ in Ref. \cite{thde5}. Therefore, the general form of $Q$, given by equation (\ref{genQ}), covers a wide range of other popular theoretical models for different choices of $b_{1}$ and $b_{2}$. Here we consider $b^2_{1}$ and $b^2_{2}$ instead of $b_{1}$ and $b_{2}$ to indicate that we only focus on the positive values of the coupling constants. As a result, $Q$ becomes positive and the energy transfers from THDE to DM which is well consistent with the Le Chatelier-Braun principle \cite{id1}. The simplicity of the functional form of $Q$ (as given in equation (\ref{genQ})), however, makes it very attractive and simple to study. Indeed, as DE and DM have not the same energy density (and hence contribution) within the universe dynamics and as we do not yet know their nature, it is reasonable to consider different contributions ($b_1\neq b_2$) for these dark components within the interaction term.\\

An possible justification of the interaction form (\ref{genQ}) may appear using  the Teleparallel Gravity (TG), based in the Weitzenb\"ock spacetime \cite{W},
 which is equivalent to General Relativity \cite{Andrade,Li}. Following Ref. \cite{Valiviita:2008iv} and in TG framework, we consider the conservation equations in presence of an interaction as $\nabla_{\alpha}T^{\alpha}_{\beta, X}=Q_{\beta, X}$
with $X=(m,D)$, and 
\begin{eqnarray}
Q_{\beta, m}=-Q_{\beta, D}=\sqrt{{\mathcal T}/6}\left(\bar{\mu}T^{\alpha}_{\alpha,m}u_{\beta,m}+\bar{\nu}T^{\alpha}_{\alpha,D}u_{\beta,
D}\right)
\end{eqnarray}
where $T^{\alpha}_{\beta, X}=p_X\delta^{\alpha}_{\beta}+(\rho_X+p_X)u^{\alpha}_X u_{\beta,X}$ for a perfect fluid and $u^{\alpha}_X=\frac{dx^{\alpha}}{\sqrt{-ds^2}}$ is the four-velocity of the fluid. Also, ${\mathcal T}$ is the scalar torsion which for the flat FRW space-time is given by ${\mathcal T}=-6H^2$ \cite{Haro}. Now, one can obtain at the background level 
\begin{eqnarray}
Q=Q_{0,c}=-Q_{0,x}=\frac{H}{\sqrt{6}}\left( \bar{\mu}\rho_c+\bar{\nu}(3w_x-1)\rho_x
\right),
\end{eqnarray}
and $\nabla_{\alpha}T^{\alpha}_{\beta, X}=-\dot{\rho}_X-3H(\rho_X+p_X)$, 
to  recover the energy conservation equations (\ref{conm}) and (\ref{conD}), one only has to take $b^{2}_{1}=\frac{\bar{\mu}}{3\sqrt{6}}$ and $b^{2}_{2}=\frac{\bar{\nu}(3w_D-1)}{3\sqrt{6}}$.\\
\par Now, taking the time derivative of equation (\ref{frd}), and by using
equations (\ref{r2}), (\ref{conm}) and (\ref{conD}), we get
\begin{eqnarray}\label{7}
\frac{\dot{H}}{H^{2}}=-\frac{3}{2}\Omega_{D}(1+\omega_{D}+r),
\end{eqnarray}
Similarly, taking the time derivative of equation (\ref{Hrho}) along with combining the result with equations (\ref{conD}) and (\ref{7}), we obtain
\begin{eqnarray}\label{w1}
\omega_{D}=\dfrac{1-\delta -\frac{b^2_{1}}{\Omega_D} + b^2_{1} - b^2_{2}}{1-(2-\delta)\Omega_{D}}.
\end{eqnarray}
For $\delta < 1$, we get $2 -\delta > 1$ which means that there exists a divergence in the evolution of $\omega_{D}$ at the redshift for which $\Omega_{D}=\frac{1}{2-\delta}$. As a result, the $\delta <1$ case is not suitable in the present work.\\
Taking the time derivative of equation (\ref{3}) and by using equations (\ref{r2}),    (\ref{7}) and (\ref{w1}), we arrive at the following equation for THDE density parameter, as  
\begin{eqnarray}\label{Omega}
\Omega_{D}^{\prime}=3(\delta-1)\Omega_{D}
\left[\dfrac{1-\Omega_{D} + {b^2_{1}\Omega_D} -b^2_{1}-{b^2_{2}\Omega_D}}{1-(2-\delta)\Omega_{D}}\right],
\end{eqnarray}
where $\Omega_{D}^{\prime}=\frac{d\Omega_D}{d(\ln a)}$.\\
Now, for simplicity, we re-expressed equation (\ref{3}) as
\begin{eqnarray}
H^{2} &=&\left(\frac{3m_{p}^{2}}{B}{\Omega_D}\right)^{\frac{1}{1-\delta}}, \nonumber \\
&=& H^2_{0} \left(\frac{\Omega_{D}}{\Omega^{0}_{D}}\right)^{\frac{1}{1-\delta}},
\end{eqnarray}
which implies,
\begin{equation}\label{eqnh}
h=\frac{H}{H_0}=\left(\frac{\Omega_{D}}{\Omega^{0}_{D}}\right)^{\frac{1}{2(1-\delta)}},
\end{equation}
where, $h$ is the normalized Hubble parameter, $\Omega^{0}_{D}$ is the present THDE density parameter and $H_{0}=\left(\frac{3m_{p}^{2}}{B}\Omega^{0}_{D}\right)^{\frac{1}{2(1-\delta)}}$, denotes the present value of $H$. Later, using equation (\ref{Omega}) along with the above equation, we try to show the evolution of $H$ for this model and will compare it with that of observational Hubble parameter data. \\
The deceleration parameter is defined as
\begin{eqnarray}\label{q}
q=-\frac{\ddot{a}}{aH^2}=-1-\frac{\dot{H}}{H^{2}},
\end{eqnarray}
which is an important cosmological parameter to investigate the expansion history of the universe. In particular, $q<0$ indicates accelerated $({\ddot{a}}>0)$ expansion phase of our universe, whereas $q>0$ indicates a decelerated phase $({\ddot{a}}<0)$. In our model, $q$ evolves as 
\begin{eqnarray}\label{q1}
q=\dfrac{(1-2\delta)\Omega_{D}+1-3{b^2_{1}}+3{b^2_{1}}\Omega_{D}-3{b^2_{2}}\Omega_{D}}{2[1-(2-\delta)\Omega_{D}]}.
\end{eqnarray}
It is well known that the jerk parameter, a dimensionless third derivative of the scale factor with respect
to cosmic time, provides a comparison between different DE models and the $\Lambda$CDM ($j=1$) model. It is given by \cite{jerk1,jerk2,jerk3}
\begin{equation}
j=\frac{\frac{d^{3}a}{dt^{3}}}{aH^3}=q(2q + 1)+(1+z)\frac{dq}{dz}.
\end{equation} 
Finally, in order to estimate the stability of the model we consider the square of sound speed given as
\begin{equation}
v^{2}_{s}=\frac{dp_{D}}{d\rho_{D}}=\omega_{D} + {\dot{\omega}}_{D} \frac{\rho_{D}}{{\dot{\rho}}_{D}}.
\end{equation}
Using then equation (\ref{conD}) along with equations (\ref{w1}) and (\ref{Omega}), the above equation can be re-expressed as
\begin{eqnarray}\label{vs2re}
v_s^2&=&\dfrac{b_1^2}{(\delta-2)\Omega_{D}(1+\Omega_{D}(\delta-2))^2}\nonumber\\&+&\dfrac{\left[1-b_2^2-\delta+b_1^2(1+\delta)+\Omega_D(\delta-1-b_1^2+b_2^2)\right]}{(1+\Omega_{D}(\delta-2))^2}.\nonumber\\
\end{eqnarray}
The sign of $v^{2}_{s}$ is important to specify the stability of background evolution. $v^{2}_{s}>0$ ($v^{2}_{s}<0$) indicates a stable (unstable) model. It is important to note here that the expressions of $q$, $\omega_D$, $\Omega_D$ and $v^{2}_{s}$ are similar to the results of \cite{tnote} for the special choice, $b_{1}=b_{2}=b$. On the otherhand, if $b_{1}=b_{2}=0$, then the equations (\ref{w1}), (\ref{q1}), (\ref{Omega}) and (\ref{vs2re}) match to the relations derived in \cite{THDE}. As discussed earlier, thus the present work is more general in the literature.  \\ 
\begin{figure}[htp]
\begin{center}
\includegraphics[width=6cm]{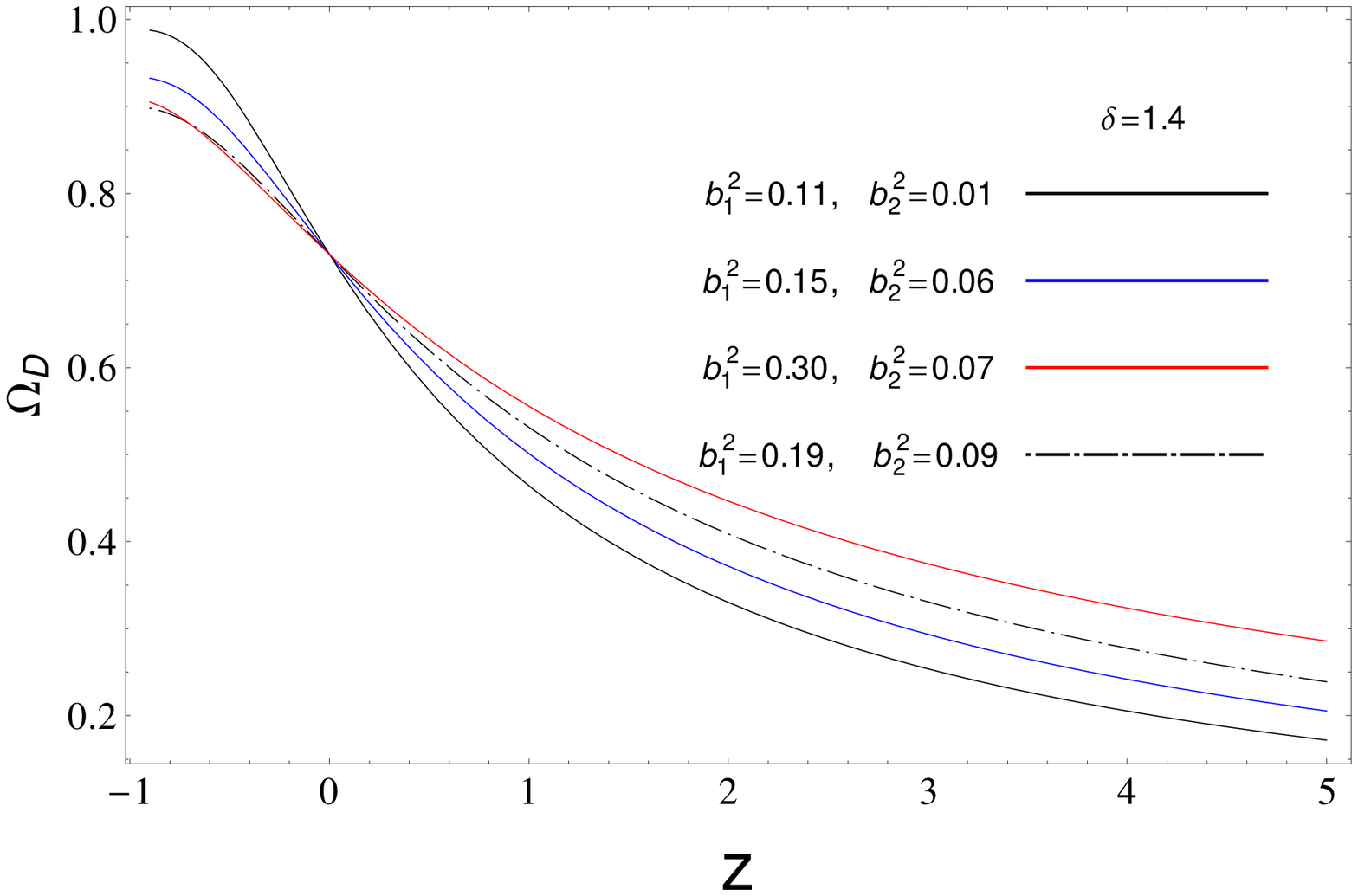}
\includegraphics[width=6cm]{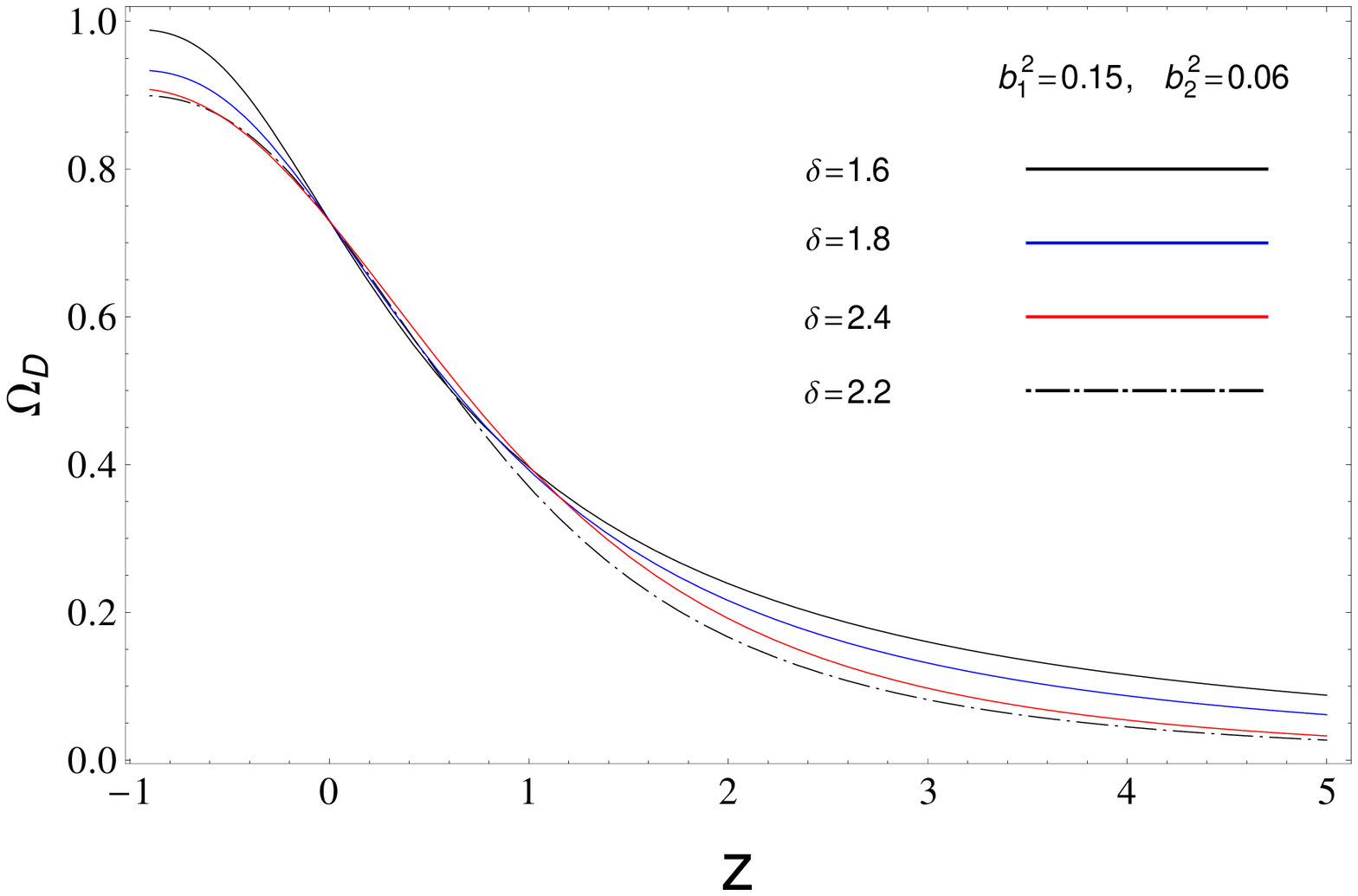}
\caption{The evolution of the THDE density parameter $\Omega_D$, as a function of $z$, is shown for the present model considering $\Omega^{0}_D=0.73$ and different values of $b^{2}_{1}$, $b^{2}_{2}$ and $\delta$, as indicated in each panel.}\label{figomegad}
\end{center}
\end{figure}
\begin{figure}[htp]
\begin{center}
\includegraphics[width=6cm]{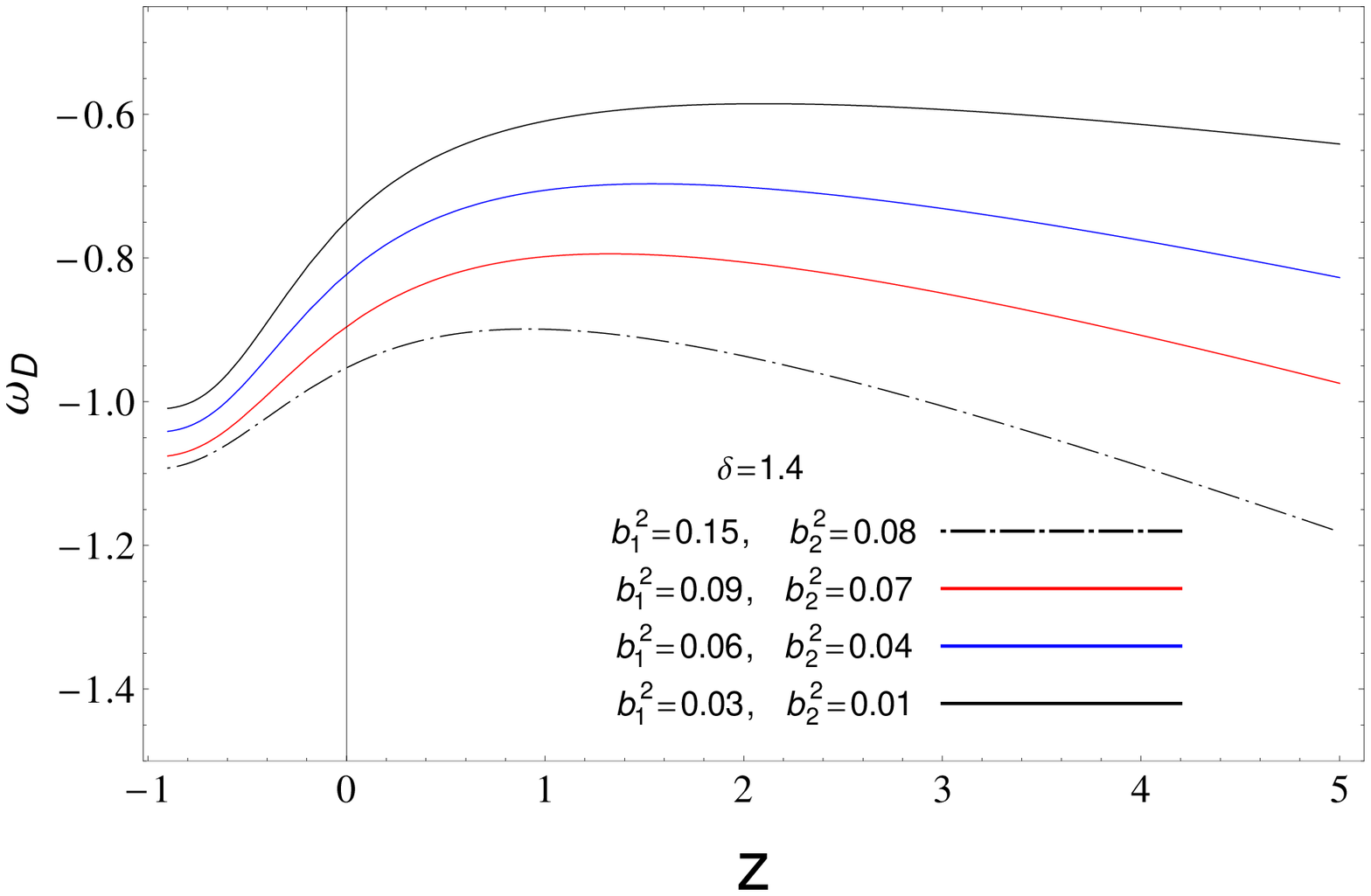}
\includegraphics[width=6cm]{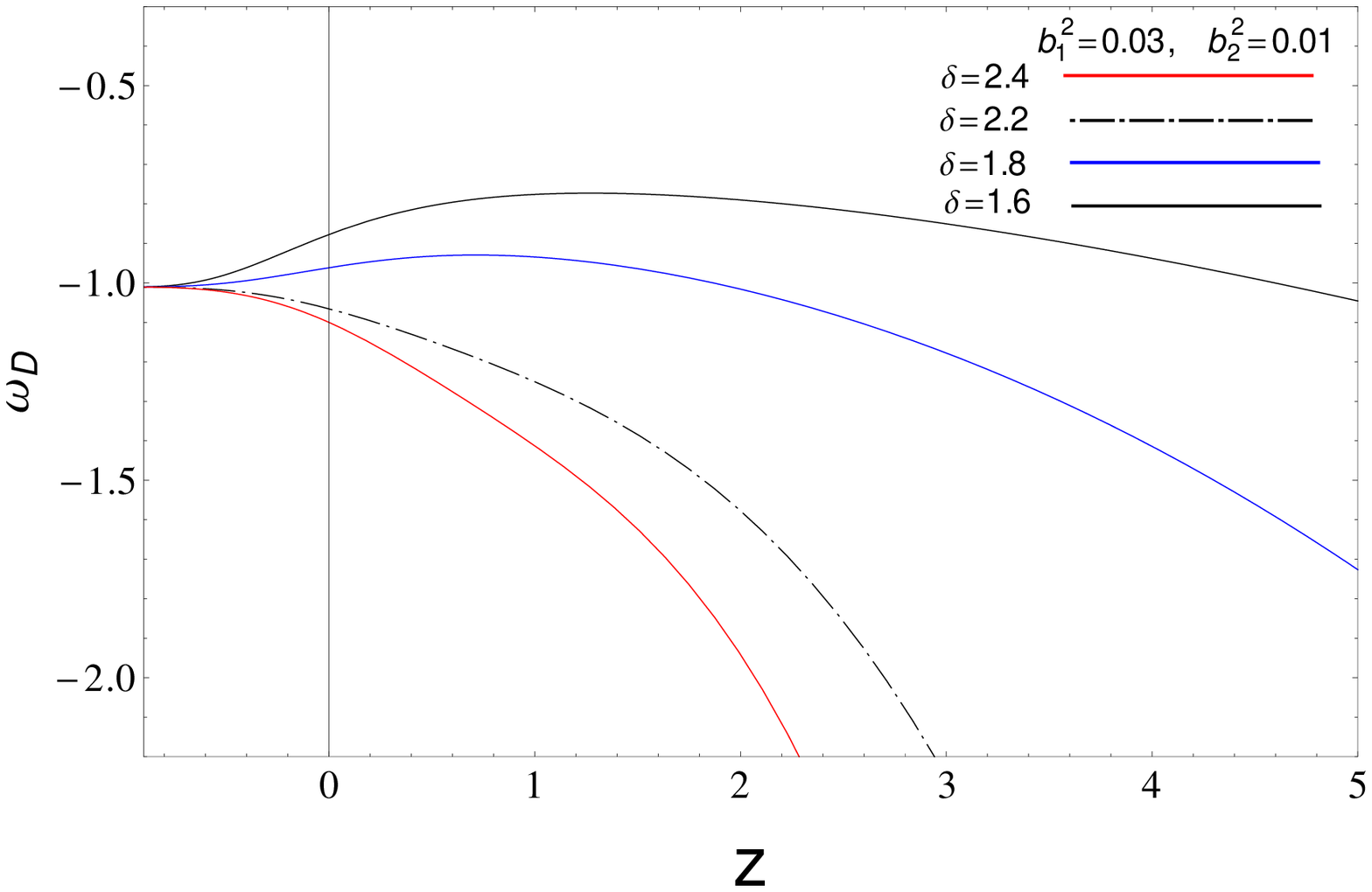}
\caption{Evolution of $\omega_D$ as a function of $z$ is shown using
$\Omega^{0}_D=0.73$ and different values of $b^{2}_{1}$, $b^{2}_{2}$ and $\delta$, as indicated in each panel.}\label{figeos}
\end{center}
\end{figure}
\begin{figure}[htp]
\begin{center}
\includegraphics[width=6cm]{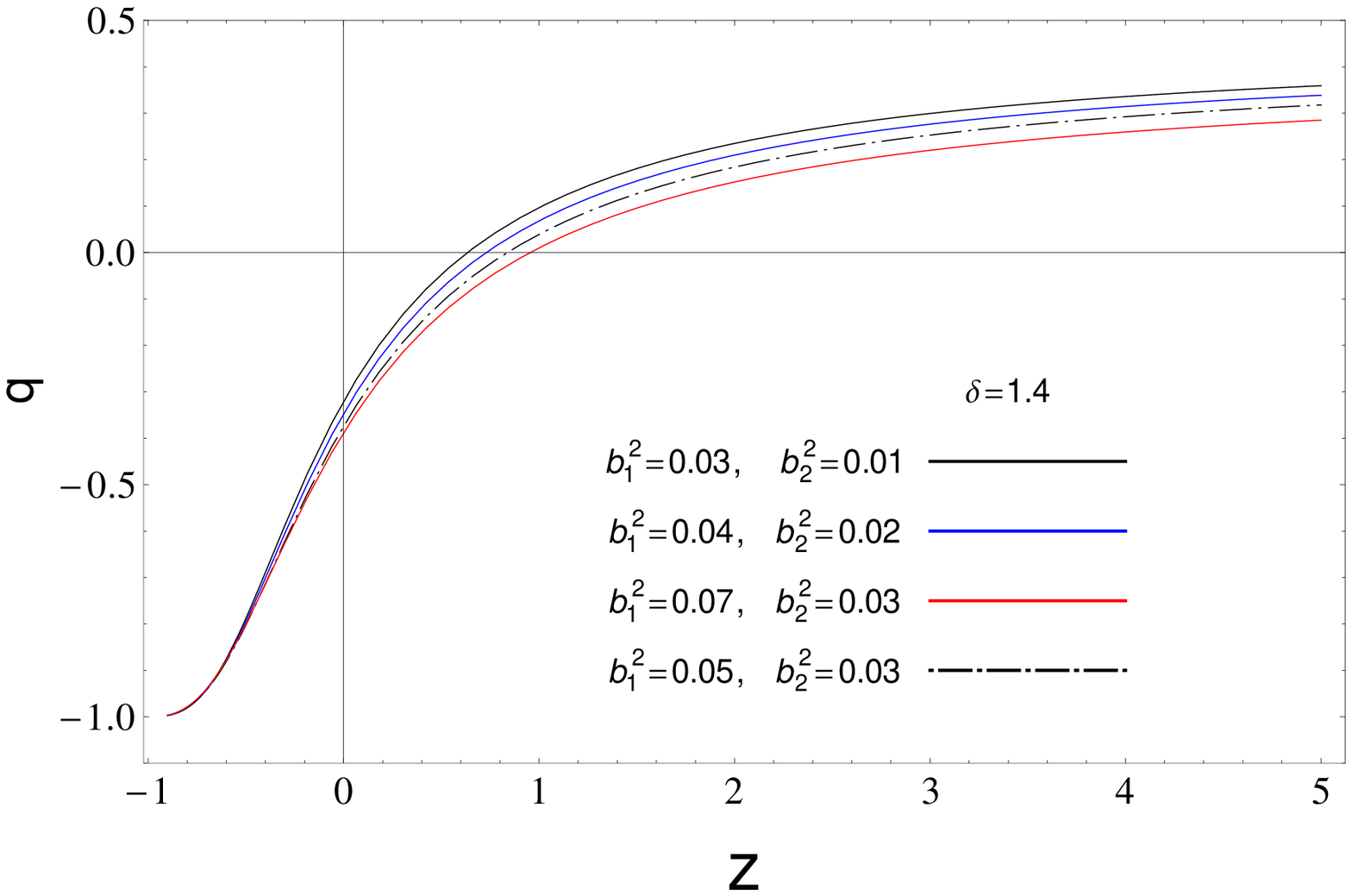}
\includegraphics[width=6cm]{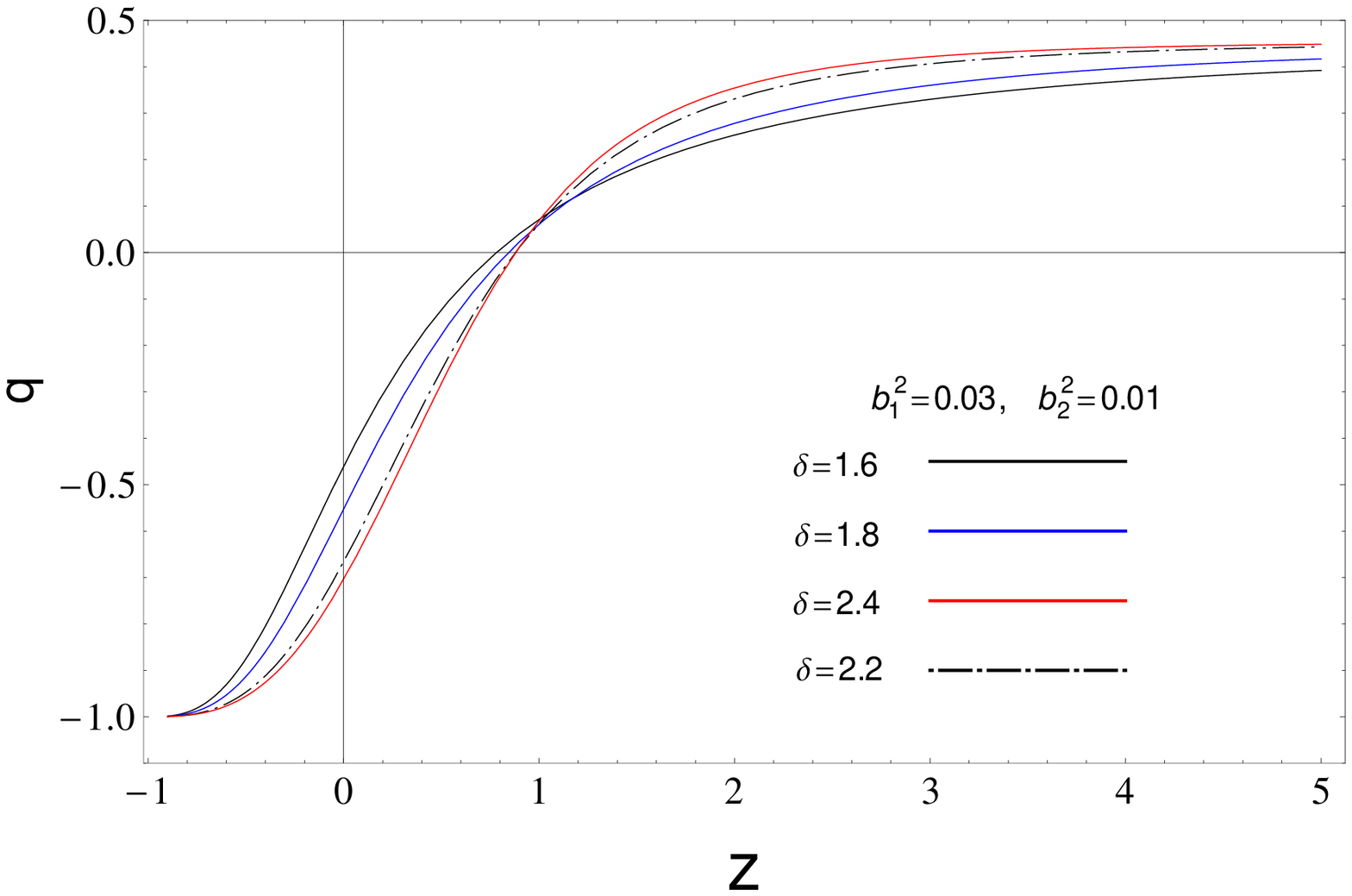}
\caption{The evolution of the deceleration parameter $q$ vs. $z$, is shown for $\Omega^{0}_D=0.73$ and different values of $b^{2}_{1}$, $b^{2}_{2}$ and $\delta$, as indicated in each panel.  Also, the horizontal line denotes $q(z)=0$.}\label{figqz}
\end{center}
\end{figure}
\begin{figure}[htp]
\begin{center}
\includegraphics[width=6cm]{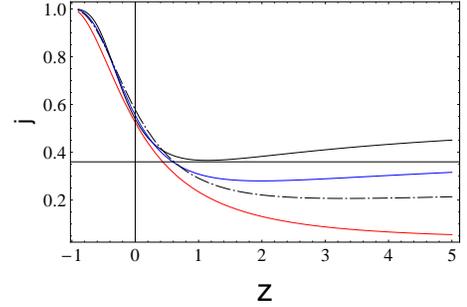}
\caption{The evolution of the cosmic jerk parameter $j(z)$ is shown for $\delta=1.4$, $\Omega^{0}_D=0.73$ and different values of $b^{2}_{1}$ and $b^{2}_{2}$, given in the upper panel of figure \ref{figomegad}.}\label{figjz}
\end{center}
\end{figure}
\begin{figure}[htp]
\begin{center}
\includegraphics[width=6cm]{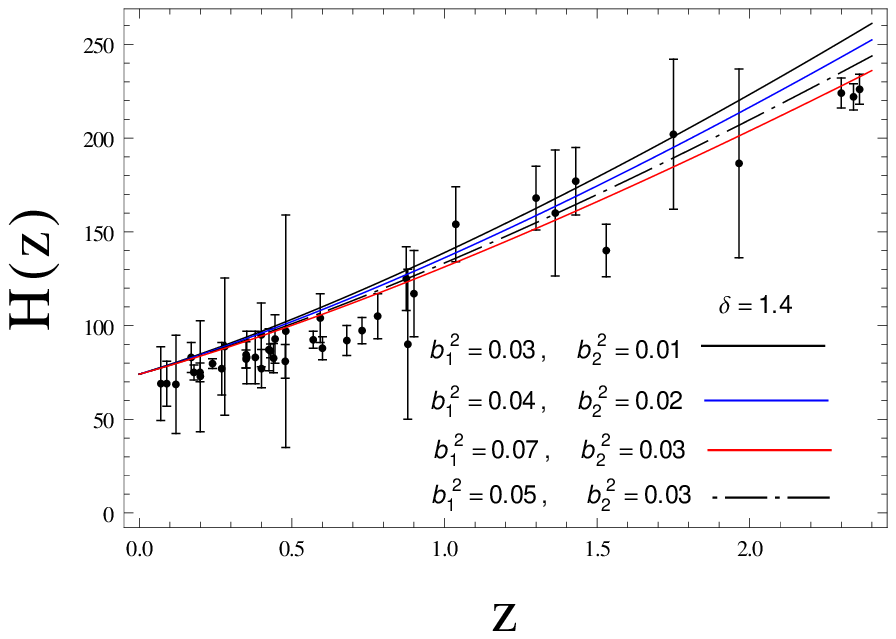}
\includegraphics[width=6cm]{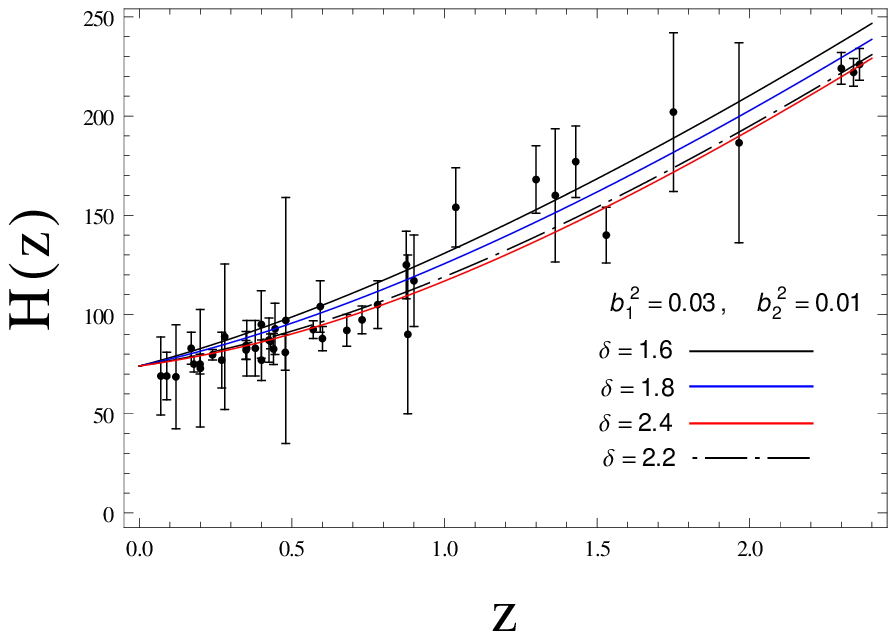}
\caption{The evolution of $H(z)$, as given in equation (\ref{eqnh}), is shown by considering $\Omega^{0}_D=0.73$ and different values of the parameters ($\delta$, $b^2_{1}$, $b^2_{2}$) , as indicated in each panel. In this plot, the black dots correspond to the $H(z)$ data consisting 41 data points with $1\sigma$ error bars \cite{hub1,hub2}. Also, the latest measurement of $H_0$ is taken from \cite{r19H0}.}\label{figh}
\end{center}
\end{figure}
\begin{figure}[htp]
\begin{center}
\includegraphics[width=6cm]{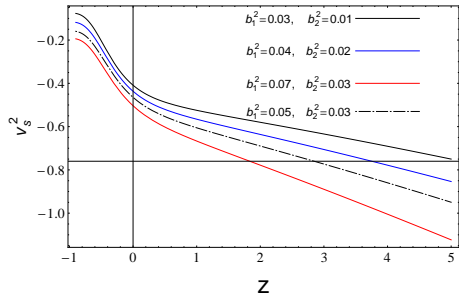}
\caption{Evolution of ${v}^{2}_{s}$ as a function of $z$ is shown for $\delta=1.4$, $\Omega^{0}_D=0.73$ and different values of $b^{2}_{1}$ and $b^{2}_{2}$, as indicated in panel.}\label{figvs1}
\end{center}
\end{figure}
\begin{figure}[htp]
\begin{center}
\includegraphics[width=6cm]{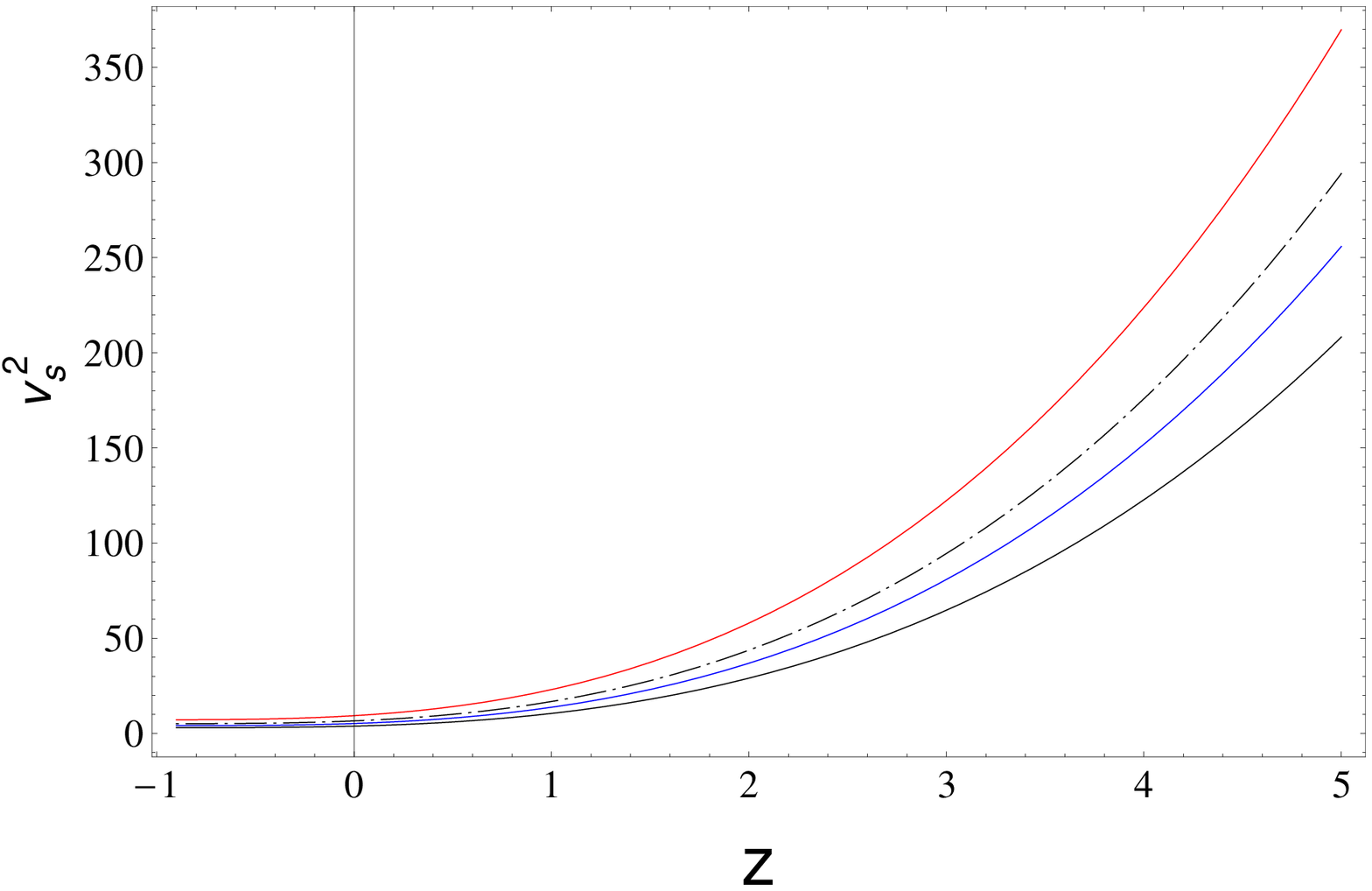}
\caption{Evolution of ${v}^{2}_{s}$ as a function of $z$ is shown for same values of $b^2_{1}$, $b^2_{2}$ and $\Omega^{0}_D$ as given in figure \ref{figvs1}. This plot is for $\delta=2.01$.}\label{figvs2}
\end{center}
\end{figure}
\par We plot the evolutionary trajectories for different cases of Tsallis parameter $\delta$ and interaction terms  $b_{1}$ and $b_{2}$. For the initial condition
$\Omega^{0}_D=0.73$, the evolutions of $\Omega_D$,
$\omega_D$, $q$, $j$ and $H$, as a function of $z$, have been plotted in
figures~\ref{figomegad},~\ref{figeos},~\ref{figqz},~\ref{figjz} and \ref{figh}, respectively. From 
the upper panel of figure \ref{figeos}, one can see that $\omega_D$ remains always in between $-1< \omega_D< -\frac{1}{3}$ at present, as expected. However, it crosses the phantom line ($\omega_{D}<-1$) in the near future as the value of the parameter pair ($b^2_{1},b^2_{2}$) increases. We also observe from the lower panel of figure \ref{figeos} that the interacting THDE model can lie in the quintessence or in the phantom regime according to the value of $\delta$. However, for the case $\delta>2$, $\omega_D$ approachs $-1$ as $z\rightarrow -1$, which means that the THDE model mimics the cosmological constant behavior in the far future.\\

 The evolution of $q(z)$, as a function of $z$, has been plotted in figure \ref{figqz}. From this figure, it is clear that our model can describe the current accelerated universe, and the transition redshift $z_t$ (i.e., $q(z_{t})=0$) from the deceleration phase to an accelerated phase occurs within the intervals [0.637,0.962] (for upper panel) and [0.776,0.889] (for lower panel), which are in good agreement with the results, $0.5<z_{t} <1$, as reported in \cite{jerk2,jerk3,zt1,zt2,zt3,zt4,zt5,zt6,zt7,zt8}. The evolution of $j(z)$ has also been plotted in figure \ref{figjz}. It is observed that $j$ stays positive and lies within (0.52-0.58) at late time, and further it tends to unity (or $\Lambda$CDM model) as $z\rightarrow -1$. This is an interesting result of the present analysis. In figure \ref{figh}, we have shown the evolution of $H$ (equation (\ref{eqnh})) for the present model and compared it with the data points for $H(z)$ (within $1\sigma$ error bars) which have been obtained from the latest compilation of 41 data points of Hubble parameter measurements ( for details, see \cite{hub1,hub2}). We observed from figure \ref{figh} that for $\delta >2$, the model reproduces the observed values of $H(z)$ quite well for each data point. Furthermore, we also checked that the nature of the evolution of $H(z)$ is hardly affected by a small change in the values of the parameters ($b^{2}_{1}$, $b^{2}_{2}$, $\delta$).\\
\par For understanding the classical stability of our model, we also plot the square of sound speed in figures~\ref{figvs1} and \ref{figvs2}. It has been found from figure \ref{figvs1} that the model is unstable ($v^{2}_{s}<0$). However, the model is stable, i.e., $v^{2}_{s}>0$, for $\delta >2$ (see figure \ref{figvs2}) and this case is not analyzed in \cite{tnote}. Thus, the stability of the interacting THDE model crucially depends on the choice of the parameter $\delta$. 
\subsection{Cosmological evolution including radiation}\label{sec-rad}
For completeness, in this subsection we extend the aforementioned scenario of
the interacting THDE model, in the case where the radiation fluid is also present. If $\rho_{r}$ is energy density of the radiation fluid, then the
Friedmann equation (\ref{frd}) becomes
\begin{eqnarray}\label{frdrad}
H^{2}=\frac{1}{3m_{p}^{2}}\left(\rho_{m}+\rho_{D}+\rho_{r}\right),
\end{eqnarray}
If we consider radiation to be decoupled from other two components (THDE and DM), then the conservation equation for radiation can be written as
\begin{equation}\label{eqconrad}
{\dot{\rho}}_{r}+4H\rho_{r}=0,
\end{equation}
Defining the following dimensionless density parameter
\begin{eqnarray}\label{omegar}
\Omega_{r}=\frac{\rho_{r}}{\rho_c}=\Omega_{r_0}(1+z)^{4},
\end{eqnarray}
together with the dimensionless density parameters $\Omega_{m}$ and $\Omega_{D}$, we find that Friedmann equation (\ref{frdrad}) can be rewritten as
\begin{eqnarray}\label{eqomrad1}
\Omega_{m}+\Omega_{D}+\Omega_{r}=1,
\end{eqnarray}
Now, differentiating the Friedmann equation (\ref{frdrad}) and using equations (\ref{conm}), (\ref{conD}), (\ref{eqconrad}) and (\ref{eqomrad1}), we obtain
\begin{eqnarray}
\frac{\dot{H}}{H^{2}}=-(1+q)=\frac{1}{2}[\Omega_{m}+\Omega_{D}(1-3\omega_{D})]-2,
\end{eqnarray}
In the case where radiation
is present, equation (\ref{w1}) now extends to
\begin{eqnarray}\label{eosthderad}
\omega_{D}=\frac{\Omega_D(1-\delta+3(2-\delta)\Omega_r)+(b_1^2-b_2^2)\Omega_D+b_1^2(\Omega_r-1)}{\Omega_D(1+(\delta-2)\Omega_D)}.\nonumber\\
\end{eqnarray}
In this case, differential equation for $\Omega_{D}$ can be obtained as
\begin{eqnarray}\label{Omegadrad}
\Omega_{D}^{\prime}=3(\delta-1)\Omega_{D}
\left[\dfrac{4-3\Omega_{m} -4\Omega_{D}- b^2_{1}\Omega_m -b^2_{2}\Omega_D}{1-(2-\delta)\Omega_{D}}\right].\nonumber\\
\end{eqnarray}
\par
\begin{figure}[htp]
\begin{center}
\includegraphics[width=6cm]{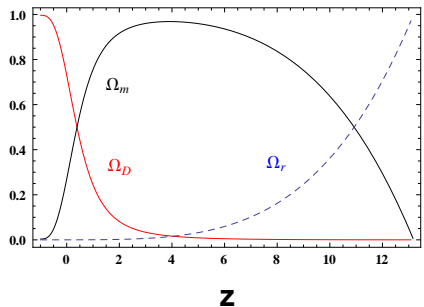}
\caption{Upper panel: The evolution of the density parameters $\Omega_D$ (red line), $\Omega_m$ (black line) and $\Omega_r$ (dashed line)  as a function of $z$ is shown for the present model considering $\Omega^{0}_D=0.73$, $\Omega_{r_0}=2.47\times10^{-5}$, $b^{2}_{1}=0.15$, $b^{2}_{2}=0.06$ and $\delta=2.1$, as indicated in each panel.}\label{figomega2}
\end{center}
\end{figure}

\begin{figure}[htp]
\begin{center}
\includegraphics[width=6cm]{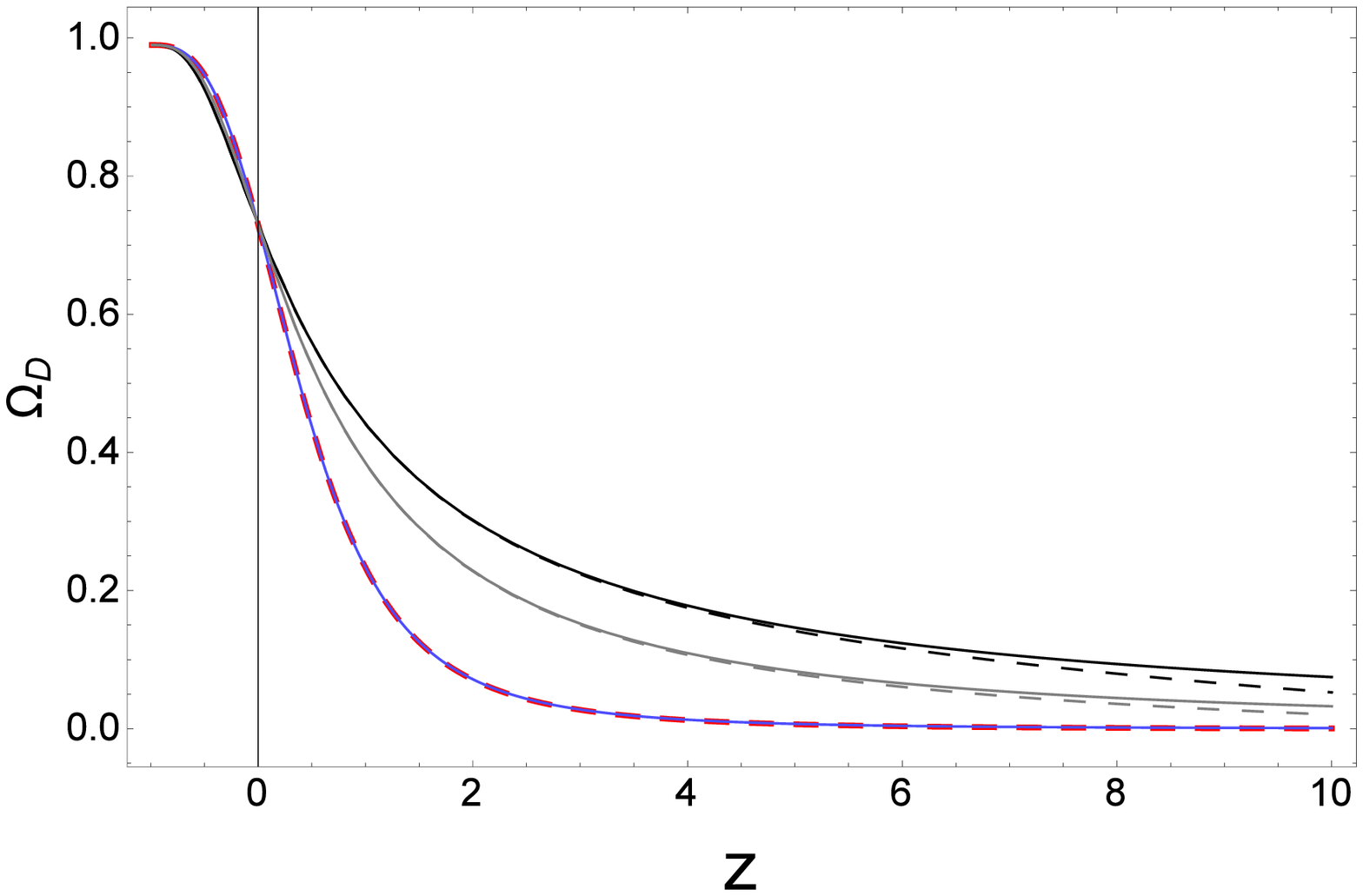}
\includegraphics[width=6cm]{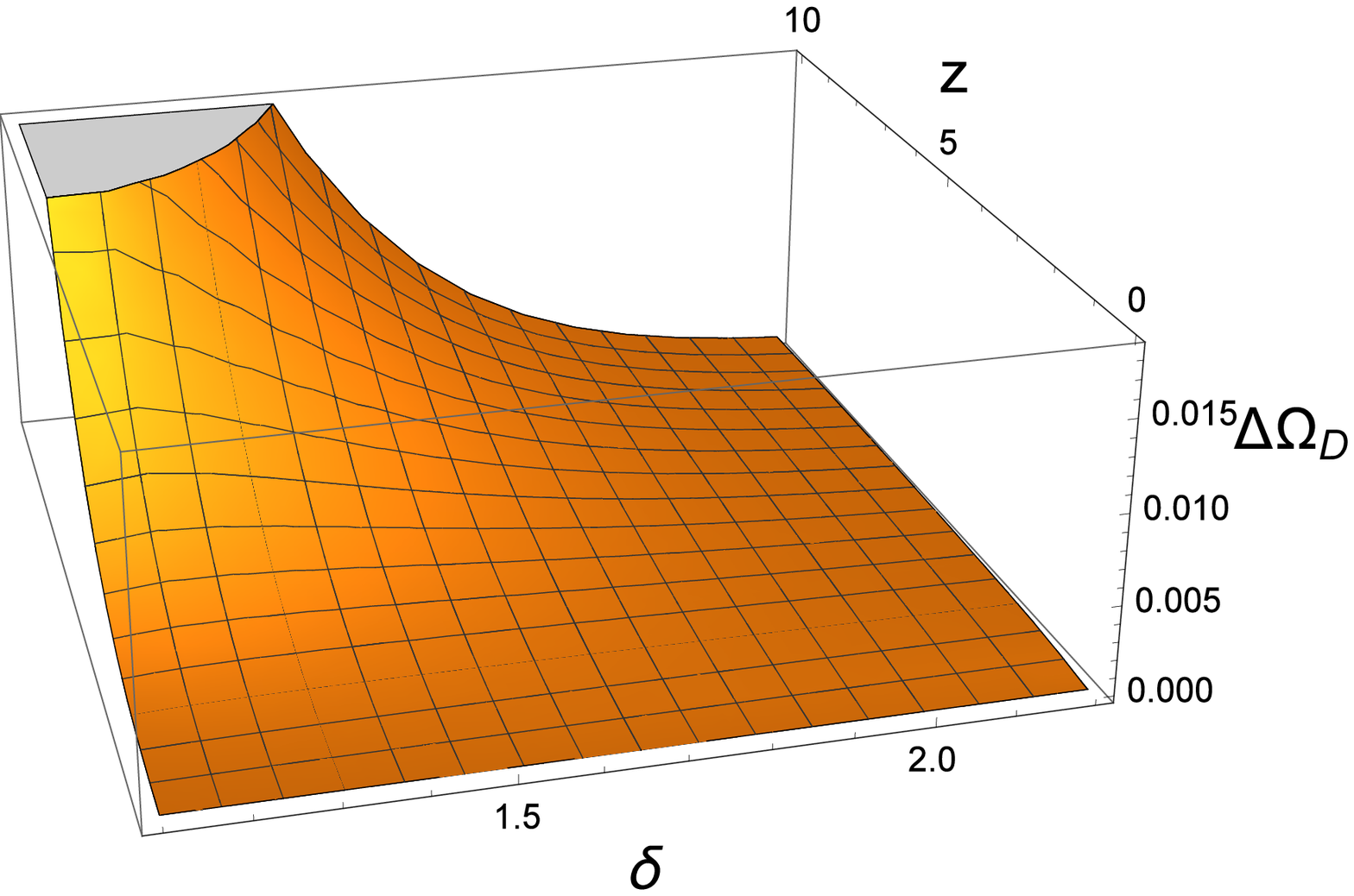}
\caption{Upper panel: The evolution of $\Omega_{D}$ versus redshift parameter $z$ for $\Omega^{0}_D=0.73$, $\Omega_{r_0}=2.47\times10^{-5}$, $b^{2}_{1}=0.03$ and $b^{2}_{2}=0.01$. The dashed (solid) black and gray curves represent evolution of $\Omega_D$ for $\delta=1.4$ and $\delta=1.55$, respectively, when radiation fluid is present (absent) and the dashed red (solid light blue) curve represents evolution of $\Omega_D$ for $\delta=2.2$ when radiation fluid is present (absent). Lower panel: Evolution of the difference between THDE density parameters with and without considering radiation as a function of redshift and $\delta$ parameter.}\label{figcomega}
\end{center}
\end{figure}

\begin{figure}[htp]
\begin{center}
\includegraphics[width=6cm]{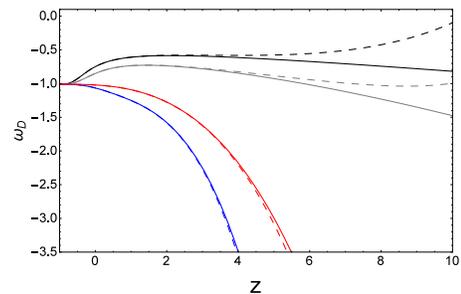}
\caption{The evolution of $\omega_D$ versus redshift parameter $z$ for $\Omega^{0}_D=0.73$, $\Omega_{r_0}=2.47\times10^{-5}$, $b^{2}_{1}=0.03$, $b^{2}_{2}=0.01$, $\delta=1.4$ (black curves), $\delta=1.55$ (gray curves), $\delta=2.0$ (red curves) and $\delta=2.2$ (blue curves). A dashed (solid) curve represents the corresponding evolution of $\Omega_D$ when radiation fluid is present (absent).}\label{figceos}
\end{center}
\end{figure}

\begin{figure}[htp]
\begin{center}
\includegraphics[width=6cm]{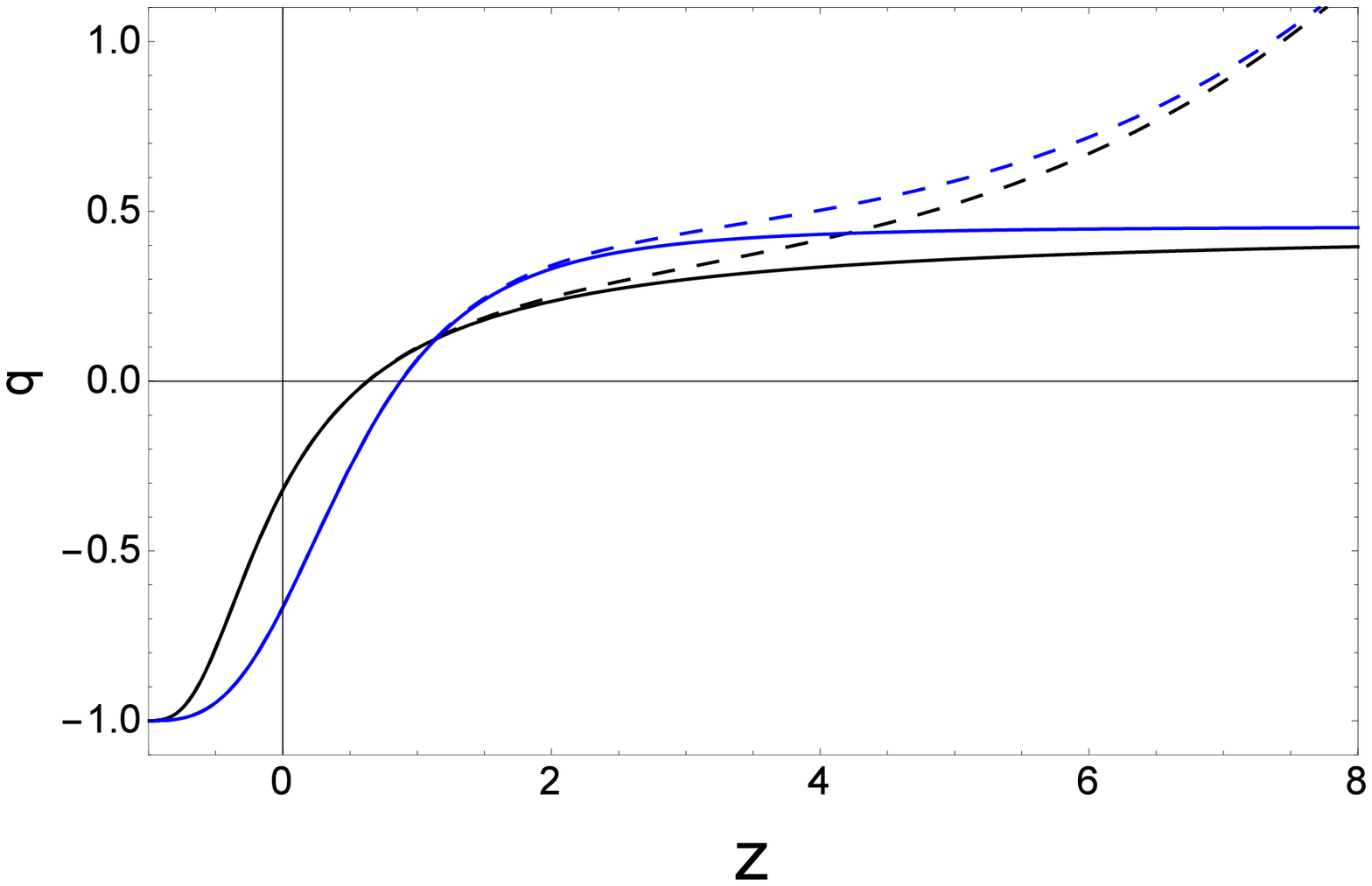}
\includegraphics[width=6cm]{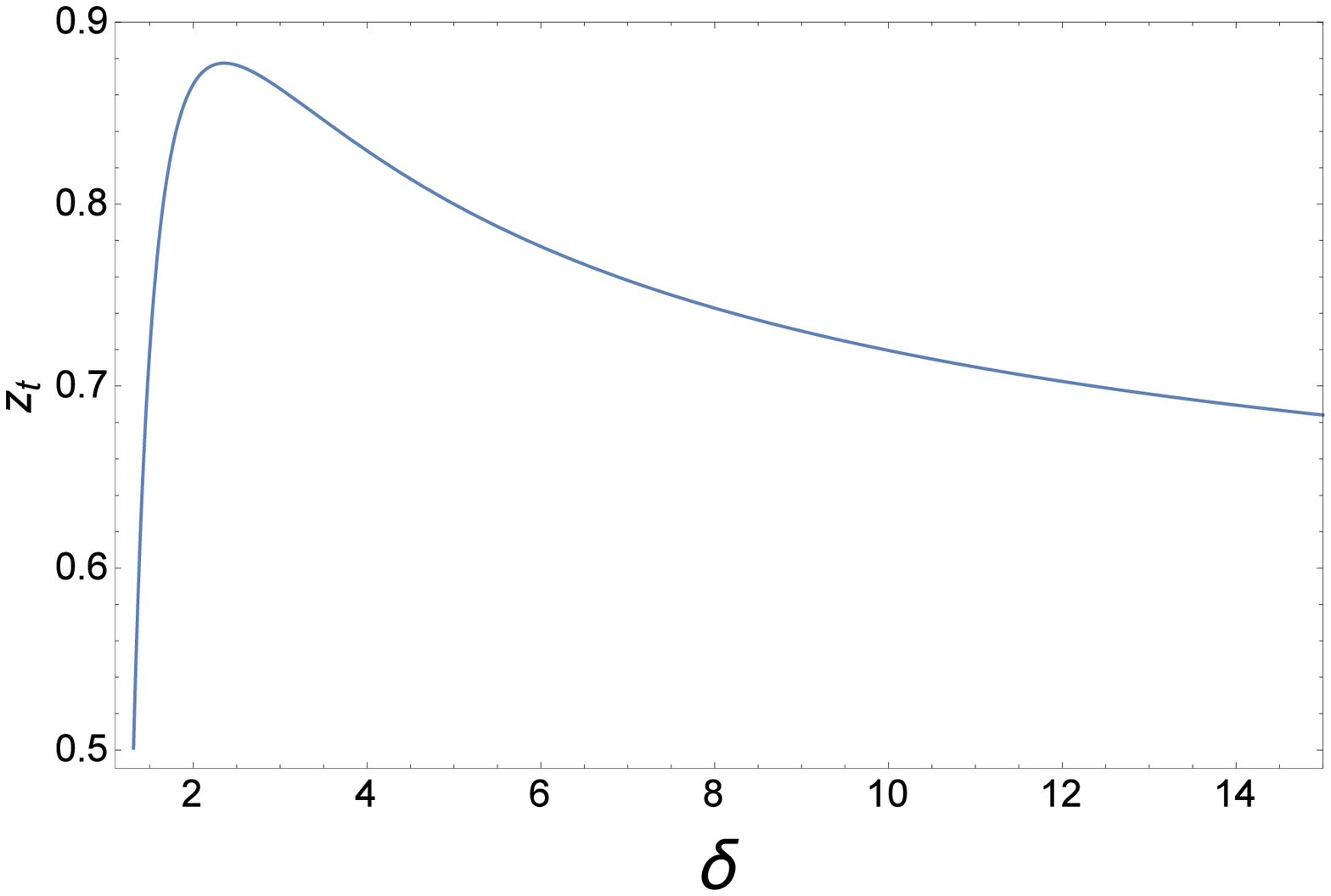}
\includegraphics[width=6cm]{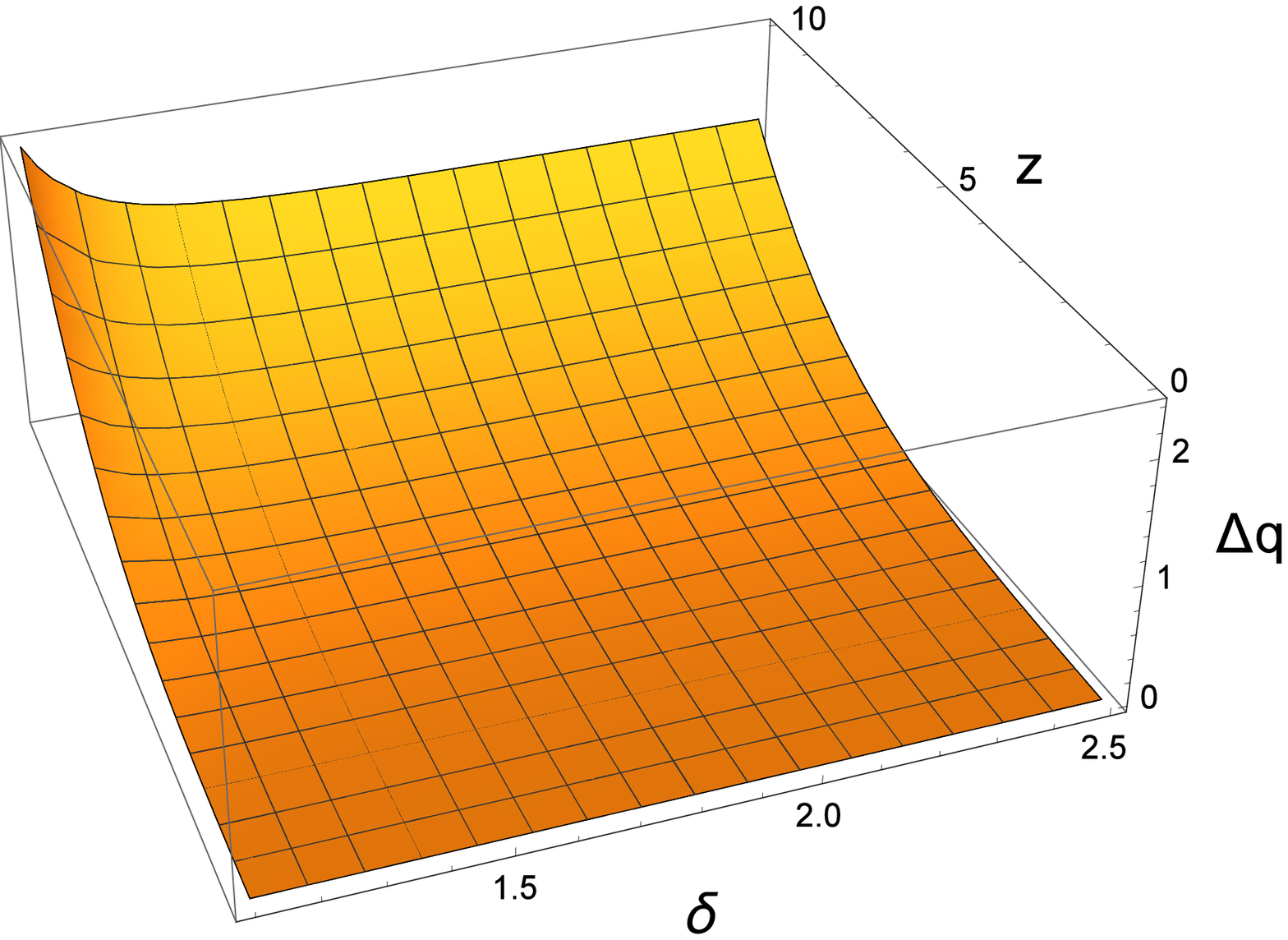}
\caption{Upper panel: The evolution of $q$ versus redshift parameter $z$ for $\Omega^{0}_D=0.73$, $\Omega_{r_0}=2.47\times10^{-5}$, $b^{2}_{1}=0.03$, $b^{2}_{2}=0.01$, $\delta=1.4$ (black curves) and $\delta=2.2$ (blue curves). A dashed (solid curve) represents the corresponding evolution of deceleration parameter when radiation fluid is present (absent). Middle panel: Transition redshift against $\delta$ parameter for the same values of model parameters as above figure. Lower panel: The behavior of the difference between deceleration parameters with and without considering radiation as a function of redshift and $\delta$ parameter.}\label{figcq}
\end{center}
\end{figure}

\begin{figure}[htp]
\begin{center}
\includegraphics[width=6cm]{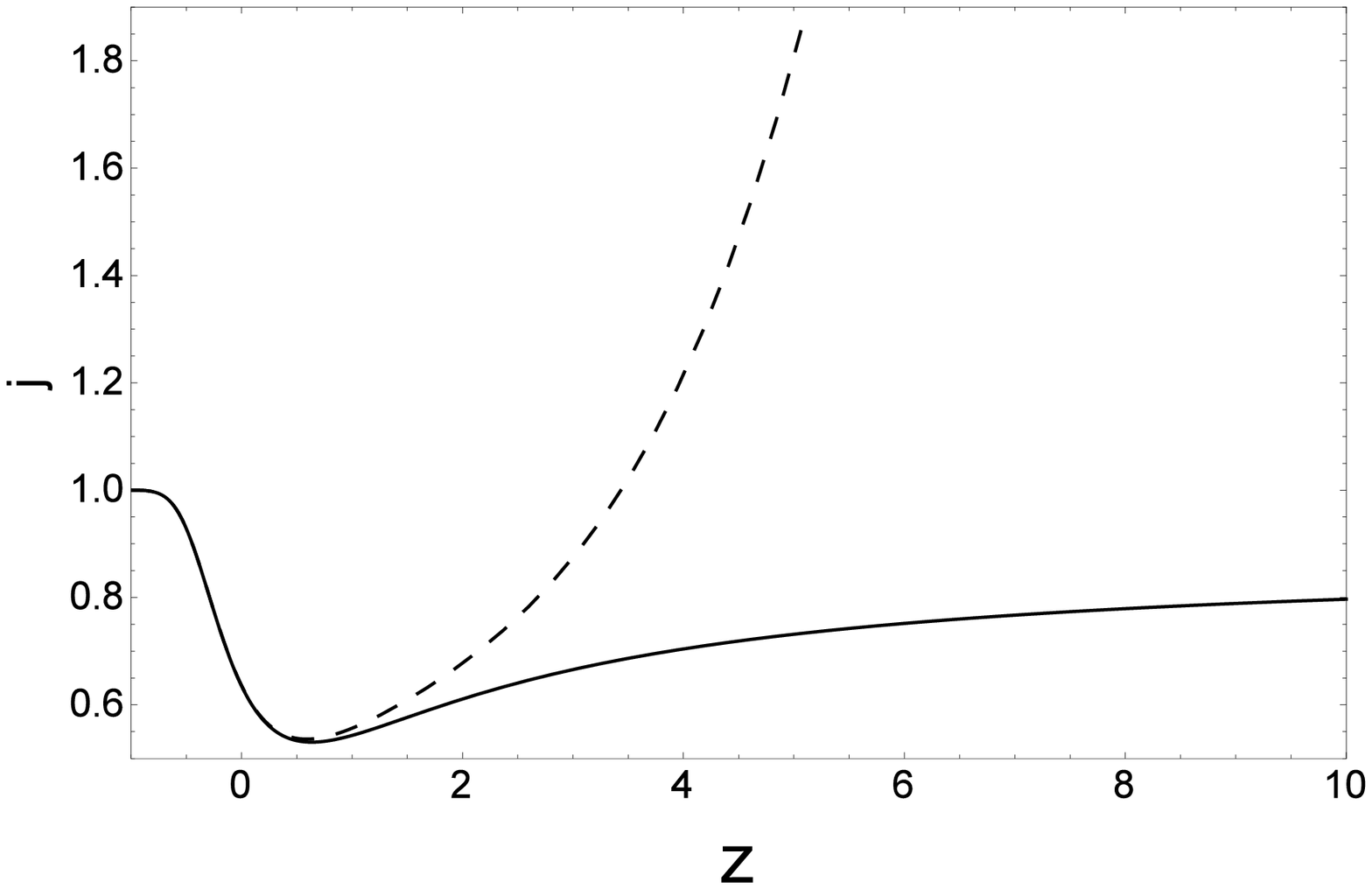}
\includegraphics[width=6cm]{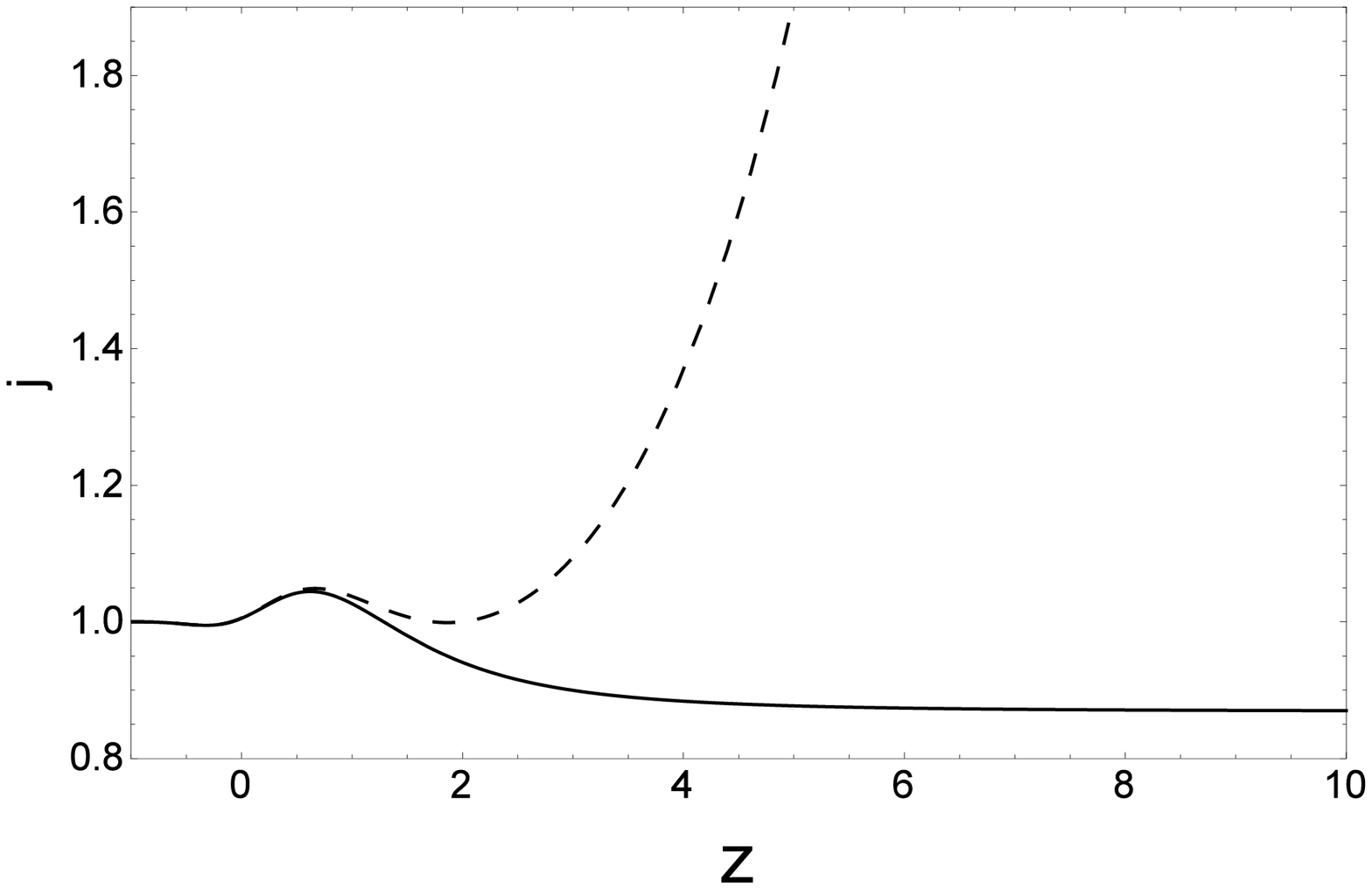}
\caption{The evolution of jerk parameter versus redshift for $\Omega^{0}_D=0.73$, $\Omega_{r_0}=2.47\times10^{-5}$, $b^{2}_{1}=0.03$, $b^{2}_{2}=0.01$, $\delta=1.4$ (upper panel) and $\delta=2.2$ (lower panel). In each panel, the dashed (solid) curve represents the corresponding evolution of $j$ when radiation fluid is present (absent).}\label{figcj}
\end{center}
\end{figure}

\begin{figure}[htp]
\begin{center}
\includegraphics[width=6cm]{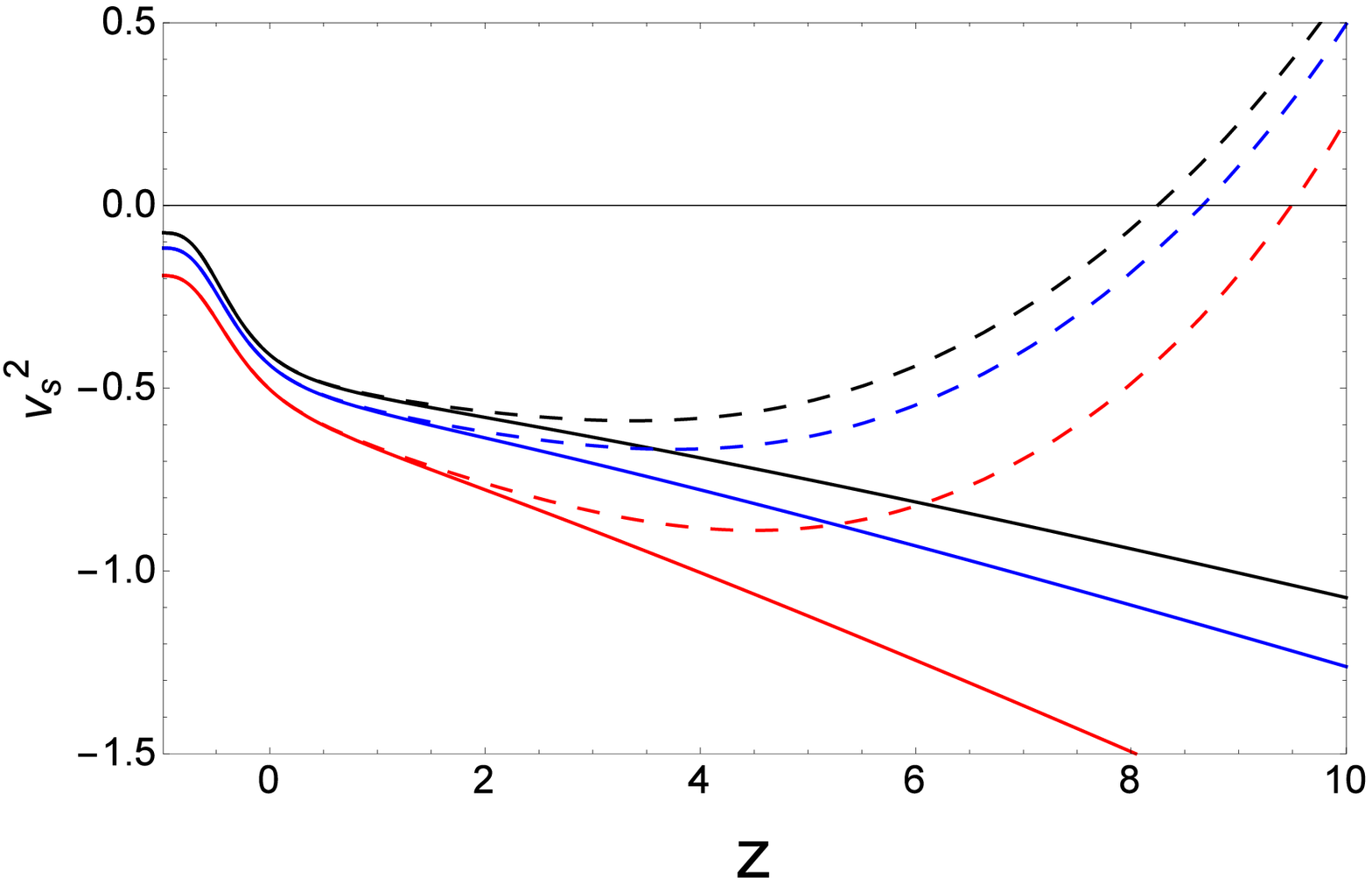}
\includegraphics[width=6cm]{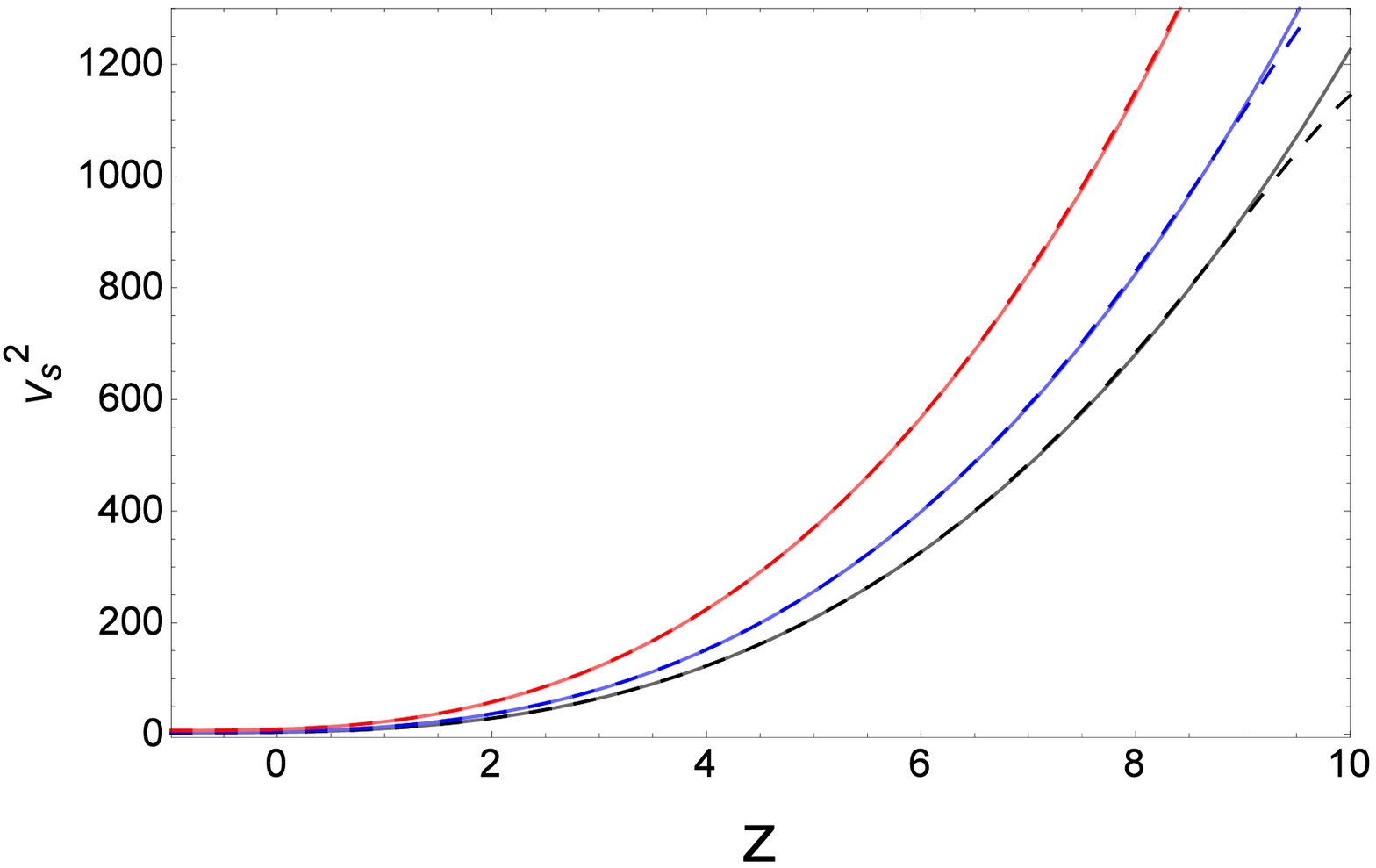}
\caption{The evolution of $v^2_{s}$ versus redshift parameter $z$ for $\Omega^{0}_D=0.73$, $\Omega_{r_0}=2.47\times10^{-5}$ and same values of $b^{2}_{1}$ and $b^{2}_{2}$ as given in Fig. (\ref{figvs1}). The upper panel is plotted for $\delta=1.4$, while the lower one is plotted for $\delta=2.01$. In each panel, the dashed (solid) curve represents the corresponding evolution of $v_s^2$ when radiation fluid is present (absent).}\label{figcvs}
\end{center}
\end{figure}

 In order to find the behavior of THDE density parameter, we solve the above equation using numerical methods. To this aim we use equations (\ref{omegar}) and (\ref{eqomrad1}) and perform numerical integration. Figure \ref{figomega2} shows the evolution of density parameters for THDE, matter and radiation. It is evident from this figure that the evolution of universe started from the radiation dominated epoch followed by the matter dominated era. The universe enters the DE dominated epoch at transition redshift $z_t$ where the deceleration parameter vanishes. This is well consistent with the usual thermal history of the universe. The upper panel in Fig. (\ref{figcomega}) presents numerical solution to Eq. (\ref{Omegadrad}) where we observe that THDE density parameter increases monotonically to unity as the universe evolves to late times ($z\rightarrow-1$). From black curves we see that the higher the redshift, the greater the difference between two curves. This is due to the fact that the effects of radiation term will be more obvious at high redshifts. However, as we increase the value of $\delta$ parameter, the magnitude of difference between these two curves ($\Delta\Omega_D=\Omega_D^{\rm withot\,rad}-\Omega_D^{\rm with\,rad}$) is reduced and thus the larger the value of $\delta$ the lesser the effects of radiation term in the evolution of THDE density parameter, see the family of gray, blue and red curves and also the behavior of $\Delta\Omega_D$ as shown in the lower panel. In Fig. (\ref{figceos}) we plotted for the equation of state parameter of THDE as given in Eq. (\ref{eosthderad}). We observe that the THDE can act as a phantom or non-phantom material during the evolution of the universe and the corresponding redshift intervals for such behavior depends crucially on $\delta$ parameter. For example, in the interval where transition occurs, i.e., $0.5<z_t<1$, the THDE can act as a fluid with phantom characteristics (see blue and red curves) and as a non-phantom fluid (see black and grays curves). We also note that as we increase the value of $\delta$ the solid and dashed curves coincide. The upper panel in Fig. (\ref{figcq}) shows the behavior of deceleration parameter against redshift where we observe that the universe evolves from an early decelerated phase, i.e., before the transition redshift, towards a late accelerated regime. We can observe that the transition redshift, $z_t$, depends on the values of $\delta$ parameter (see the middle panel) in such a way that, as this parameter increases, $z_t$ increases too, until reaching a maximum value after which for larger values of $\delta$, the transition redshift decreases monotonically to a certain nonzero value. In the lower panel we have plotted for differences between deceleration parameter with and without considering radiation term ($\Delta q(z,\delta)=q(z)^{\rm without\,rad}-q(z)^{\rm with\,rad}$) in terms of the  redshift and $\delta$ parameter. We observe that the greater the values of $\delta$ parameter, the smaller the values that $\Delta q$ assumes. It also decreases (increases) at later (earlier) times. Fig. (\ref{figcj}) shows the behavior of jerk parameter against redshift where we observe that this parameter stays positive for both cases, i.e., with and without considering radiation term and we have $j\rightarrow1$ at late times. However, at higher redshifts, the contribution due to radiation terms dominate and it is then expected that the two curves deviate as the redshift increases. Finally, Fig. (\ref{figcvs}) presents the behavior of square of sound speed against redshift where we observe that for $\delta>2$ we have $v_s^2>0$ and thus the model is classically stable, see the lower panel. Also the radiation term becomes important at high redshift and the two curves depart from each other as the redshift increases. However, as the upper panel shows, when the effects of radiation are taken into account, the model can be classically stable at early times (higher redshifts) and becomes unstable as the universe evolves to later times.
\section{Thermodynamics of interacting THDE}\label{sec-thermo}
In this section, we derive the rate of change of the total entropy and then examine the validity of generalized second law of thermodynamics. It is well known that thermodynamical analysis of the gravity theory is an exciting research topic in the cosmological context and the thermodynamical properties which hold for a black hole are equally valid for a cosmological horizon \cite{Bekenstein,Hawking,gibbons, jacobson, paddy,bak-rey, horizon-temperature-1, horizon-temperature-2,  horizon-temperature-3,  horizon-temperature-4, horizon-temperature-5,jamil,eqtemp}. In addition, the first law of thermodynamics which holds in a black hole horizon can also be derived from the first Friedmann equation in the FRW universe when the universe is bounded by an apparent horizon. This provides  well motivation to select the apparent horizon as the cosmological horizon in order to examine the thermodynamic properties of any cosmological model. Motivated by the above arguments, here, we have considered the universe as a thermodynamic system that is bounded by the cosmological apparent horizon with the radius \cite{bak-rey}
\begin{equation}
r_h= \left(H^2+ \frac{k}{a^2}  \right)^{-1/2}.
\end{equation}  
For a spatially flat universe ($k= 0$), the above equation immediately give
\begin{equation}
r_h = \frac{1}{H},
\end{equation} 
which is the Hubble horizon.\\ 
If we consider $S_f$ and $S_h$  are the entropy of the fluid and the entropy of the horizon containing the fluid, then the total entropy ($S$) of the system can be expressed as
\begin{equation}
S= S_f+ S_h.
\end{equation}
According to the laws of thermodynamics, like any isolated macroscopic system, then $S$ should satisfy the following relations
\begin{equation}
\dot{S}=\frac{dS}{dt}~\ge ~0~~~~~{\rm and}~~~~~\ddot{S}=\frac{d^{2}S}{dt^{2}}<0.
\end{equation}
In this context, it is important to mention that the generalized second law (GSL) of thermodynamics and thermodynamic equilibrium (TE) refer to the inequalities $\dot{S}~\ge ~0$ and $\ddot{S}<0$ respectively. Furthermore, the GSL should be true throughout the evolution of the universe, while the TE should hold at least during the final phases of its evolution. We shall now  examine the validity
of GSL of thermodynamics in the present context.\\
\par Now, the entropy of the horizon $S_h$ can be derived as \cite{tsahde,THDE}, 
\begin{eqnarray}\label{sp-thermo2}
S_h= \gamma A^{\delta} = \gamma (4\pi)^{\delta} r_h^{2\delta},
\end{eqnarray} 
where, $\gamma$ is an unknown constant and $\delta$ denotes the non-additivity parameter, as mentioned in equation (\ref{Trho}). Here, $A= 4 \pi r_h^2$ and $r_h$ are the surface area and radius of the apparent horizon respectively. It is important to note here that for $\delta=1$ and $\gamma=\frac{1}{4G}$ (in units where $\hslash=k_{B}=c=1$), the expression (\ref{sp-thermo2}) gives the usual Bekenstein entropy \cite{jacobson,horizon-temperature-4}. Also, the temperture of the apparent horizon is given by the relation \cite{horizon-temperature-4}
\begin{equation}
T_h = \frac{1}{2 \pi r_h}.
\end{equation}
\par As previously mentioned, we considered the THDE, DM and radiation as the components in the energy budget, so we can write
\begin{equation}
S_f =S_D + S_m + S_r,
\end{equation}
where, $S_D$, $S_m$ and $S_r$ represent the entropies of the THDE, DM and radiation respectively, and $T$ is the temperture of the composite matter inside the horizon. Therefore, the first law of thermodynamics ($ TdS = dE + p dV$) can be written for the individual matter contents in the following form
\begin{align}
T d S_D &= dE_D + p_D dV\label{sp-thermo5},\\
T dS_m &= dE_m + p_m dV  = dE_m\label{sp-thermo4},\\
T d S_r &= dE_r + p_r dV\label{sp-thermo5rad},
\end{align}
where $V= \frac{4}{3}\pi r_h^3$, is the horizon volume. Also, $E_D= \frac{4}{3}\pi r_h^3 \rho_D$, $E_m=\frac{4}{3} \pi r_h^3 \rho_m$ and $E_r=\frac{4}{3} \pi r_h^3 \rho_r$ represent the internal energies of the THDE, DM ($p_{m}=0$) and radiation ($p_{r}=\frac{1}{3}\rho_r$) respectively. Now, differentiating equations (\ref{sp-thermo2}), (\ref{sp-thermo5}), (\ref{sp-thermo4}) and (\ref{sp-thermo5rad})  with respect to time, we obtain
\begin{eqnarray}\label{eqsda}
{\dot{S}_h}&=&2\gamma \delta (4\pi)^{\delta} r^{2\delta -1}_h \dot{r}_h , \nonumber \\
{\dot{S}_D}&=&\frac{4 \, \pi \, p_D \, r_h^2 \,\dot{r}_h + \dot{E}_D}{T} , \nonumber \\
{\dot{S}_m}&=&\frac{\dot{E}_m}{T}, \nonumber \\
{\dot{S}_r}&=&\frac{4 \, \pi \, p_r \, r_h^2 \,\dot{r}_h + \dot{E}_r}{T} ,
\end{eqnarray}
Lastly, a crucial asumption in this context is that the fluid temperture $T$ should be equal to that of the horizon temperture $T_h$, otherwise the
energy flow would deform this geometry (for details, see \cite{id1,horizon-temperature-4,horizon-temperature-5,jamil,eqtemp}). Along with this assumption and   using equation (\ref{eqsda}),  we arrive at the expression 
\begin{eqnarray}\label{gslt}
\dot{S}&=& {\dot{S}}_{D}+{\dot{S}}_{m}+{\dot{S}}_{r}+{\dot{S}}_{h}\nonumber \\
&=&\frac{2\pi}{G}\frac{\dot{H}}{H^5}\left[\dot{H}+H^{2}-\frac{\gamma G\delta}{\pi}(4\pi)^{\delta}H^{4-2\delta}\right]. 
\end{eqnarray}
It deserves to mention here that the above result holds independently of the interaction form $Q$. For $\delta=1$ and $\gamma=\frac{1}{4G}$, the expression (\ref{gslt}) reduces to the case of usual Bekenstein entropy given by
\begin{equation}\label{eq-bhspstot}
{\dot{S}}= \frac{2\pi}{G}\frac{{\dot{H}}^2}{H^5},
\end{equation}
which is always positive definite irrespective of the functional forms of $H$. This fact proves the validity of the GSL of thermodynamics at all cosmological times. In fact, the relation (\ref{eq-bhspstot}), in units of $8\pi G=1$, has already been established in the context of interacting DE, where DE, DM and radiation are inteacting with each other \cite{jamil}.\\
\par However, it evident from equation (\ref{gslt}) that for the present model, the GSL will be valid if either (i) ${\dot{H}}>0$ and $\left[\dot{H}+H^{2}-\frac{\gamma G\delta}{\pi}(4\pi)^{\delta}H^{4-2\delta}\right]>0$ or (ii) ${\dot{H}}<0$ and $\left[\dot{H}+H^{2}-\frac{\gamma G\delta}{\pi}(4\pi)^{\delta}H^{4-2\delta}\right]<0$. Hence, the total
entropy $S$ is not necessarily an increasing function of
time, and the GSL of thermodynamics may be violated, depending on evolution of the universe.
\section{Conclusions}\label{conclusion}
In this paper, we have studied an accelerating cosmological model for the present universe which is filled with DM and THDE. The DM is assumed to be interact with the THDE whose IR cut-off scale is set by the Hubble length. As already discussed in section \ref{sec2}, the functional form of $Q$ is chosen in such a way that it reproduces well known and most used interactions in the literature for some specific values of the model parameters $b_{1}$ and $b_{2}$ \cite{tnote,thde5,ig1,id1,im1,im2}.\\ 
\par In our setups, the behavior of various quantities, e.g., $\Omega_{D}$, $\omega_{D}$, $q$, $j$, $H$ and $v^2_{s}$ have been studied during the cosmic evolution. We have also found that the case $\delta >2$ can produce suitable behavior for the parameters $\Omega_{D}$, $\omega_{D}$, $q$ and $v^2_{s}$, but this case is not analyzed in \cite{tnote}. It is observed that the Tsallis parameter $\delta$ significantly affects the THDE equation of state parameter, and according to its value it can lead it to lie in the phantom regime or in the quintessence regime during the evolution, before it asymptotically stabilizes in the cosmological constant value at future. The evolution of $q$ shows that the universe is decelerating at early epoch and accelerating at present epoch. This explains both the observed growth of structures at the early times and the late time cosmic acceleration measurements. Also, the transition between the DM era and the THDE era takes place within the redshift interval [0.637,0.962], which are in good compatibility with several recent studies \cite{jerk2,jerk3,zt1,zt2,zt3,zt4,zt5,zt6,zt7,zt8}. It is also observed that $j$ stays positive and approaches to the $\Lambda$CDM ($j=1$) model as $z\rightarrow -1$. Further, we studied the thermodynamic nature of the universe for this model. The basic motivation was to verify whether our model fulfills the thermodynamical requirements of the expanding universe. Our study shows that for Bekenstein entropy ($\delta=1$), the GSL of thermodynamics is always satisfies. However, the GSL of thermodynamics may be violated for $\delta \neq 1$, depending on evolution of the universe. \\ 
\par
\begin{figure}[htp]
	\begin{center}
		\includegraphics[width=7cm]{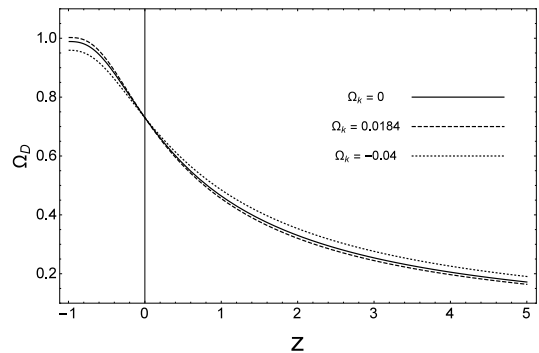}
		\caption{The evolution of the THDE density parameter $\Omega_D$, as a function of $z$, for $\Omega^{0}_D=0.73$, $b^{2}_{1}=0.011$, $b^{2}_{2}=0.01$ and $\delta=1.4$. The values of curvature density parameter has been extracted from the data provided in~\cite{Dio2016}.}\label{fignonflat}
	\end{center}
\end{figure}

 Furthermore, we noticed that the stability of our model against small perturbations during the cosmic evolution, crucially depends on the choice of the parameter $\delta$ (see figures \ref{figvs1} \& \ref{figvs2}). Therefore, we conclude that for the deep understanding of behavior of interacting THDE, more investigations should be done. In this context, it deserves to mention here that Sharma et al. \cite{newref1} recently explored an interacting THDE model (with $b^{2}_{1}=b^{2}_{2}=b^{2}$) in the framework of a non-flat universe and also by considering apparent horizon as IR cutoff from the statefinder and $\omega_{D}-{\omega}^{\prime}_{D}$ pair viewpoint. They showed the evolution of $q$, for different Tsallis parameter $\delta$ and $b^2$ and also distinct spatial curvature contributions corresponding to the flat, closed and open universes, respectively. Moreover, they found the values of the transition redshift for any spatial curvature, however the difference between them is minor and the transition from decelerated phase to accelerated phase is fully consistent with the observational data. In another recent work, Papagiannopoulos et al. \cite{newref2} studied the dynamical properties of a large body of varying vacuum cosmologies for which DM interacts with vacuum. In particular, they investigated the existence and the stability of cosmological solutions by performing the critical point analysis. Later, Panotopoulos et al. \cite{newref2a} studied three interacting dark energy models utilizing dynamical system tools and statefinder analysis, which have the potential to discriminate between various dark energy models. Moreover, the study of THDE model in a non-flat universe within the framework of dynamical Chern-Simon modified gravity has been done in~\cite{Jawad2019} where the authors have found compatibility of Hubble parameter and cosmological evolution of deceleration parameter with present day observational data. Work along this line has been also extended to modified gravity theories such as Brans-Dicke theory~\cite{Ghaffari2018}. In this regard if we consider a non-flat FRW spacetime, Eq.(\ref{Omega}) will be modified as follows
\begin{equation}
\Omega^\prime_{D}=(\delta-1)\Omega_{D}\left[\frac{3+3b_1^2(\Omega_{D}-1)-3\Omega_{D}(1+b_2^2)+2\Omega_k}{1+\Omega_D(\delta-2)}\right],
\end{equation}
where $\Omega_k=k/a^2H^2$ is the density parameter of spatial curvature. In Fig.(\ref{fignonflat}) we have plotted for the behavior of THDE density parameter against redshift where we observe that the overall behavior of this parameter in flat case is close to the non-flat cases. However, for $z>0$ and $\Omega_k>0$, the THDE density parameter assumes lesser values in comparison to the flat case while $\Omega_D$ for negative spatial curvature is greater than the flat one. As the universe evolves to the present time, the position of the curves with negative and positive spatial curvature changes and at the late times, the THDE density parameter for $\Omega_k>0$ dominates the case with $\Omega_k<0$ and the flat one.
\par It is therefore reasonable to examine the results of present study by following the methods as presented in~\cite{newref1,newref2,newref2a}. This also helps us to find out possible differences between flat and non-flat cases in a more concrete way. In a follow-up study, we would like to study the model by considering other IR cutoffs and some non-linear interaction between the dark sectors, which may modify the properties of THDE. 
\section{Acknowledgments}
The authors are thankful to the anonymous referee whose useful suggestions have improved the quality of the paper. The work of KB was supported in part by the JSPS KAKENHI Grant Number JP 25800136 and Competitive Research Funds for Fukushima University Faculty (19RI017). 

\end{document}